\documentclass[useAMS,twocolumn,usenatbib]{mnras}

\usepackage{graphicx,enumitem,color,hyperref}

\usepackage{amssymb,amsmath,xtab}

\newcommand{\lya}{Lyman-$\alpha$ }

\def\HI{\hbox{H~$\scriptstyle\rm I\ $}}
\def\HII{\hbox{H~$\scriptstyle\rm II\ $}}
\def\nHI{{\rm HI}}
\def\nH{{\rm H}}
\def\nHII{{\rm HII}}
\def\nHe{{\rm He}}
\def\nHeI{{\rm HeI}}
\def\nHeII{{\rm HeII}}
\def\nHeIII{{\rm HeIII}}

\def\HeI{\hbox{He~$\scriptstyle\rm I\ $}}
\def\HeII{\hbox{He~$\scriptstyle\rm II\ $}}

\def\HeIII{\hbox{He~$\scriptstyle\rm III\ $}}

\def\lumdens{\,{\rm erg\,s^{-1}\,Mpc^{-3}\,Hz^{-1}}}

\def\sfrd{\,{\rm M_\odot\,yr^{-1}\,Mpc^{-3}}}

\def\Lya{Ly$\alpha\ $}

\def\spose#1{\hbox to 0pt{#1\hss}}
\def\lta{\mathrel{\spose{\lower 3pt\hbox{$\mathchar"218$}}
     \raise 2.0pt\hbox{$\mathchar"13C$}}}
\def\gta{\mathrel{\spose{\lower 3pt\hbox{$\mathchar"218$}}
     \raise 2.0pt\hbox{$\mathchar"13E$}}}

\hypersetup{draft} 
\begin{document}

\title[Consistent modelling of UV background and IGM]{Consistent modelling of the meta-galactic UV background and the thermal/ionization history of the intergalactic medium}
\author[E. Puchwein, F. Haardt, M. Haehnelt, P. Madau] {Ewald Puchwein$^{1}$, Francesco Haardt$^{2,3}$, Martin G. Haehnelt$^{1}$, Piero Madau$^{4,5}$\vspace{2mm}
\\$^1$Institute of Astronomy and Kavli Institute for Cosmology, University of Cambridge, Madingley Road, Cambridge CB3 0HA, UK
\\$^2$DiSAT, Universit\`a dell'Insubria, Via Valleggio 11, I-22100 Como, Italy
\\$^3$INFN, Sezione di Milano-Bicocca, Piazza delle Scienze 3, I-20123 Milano, Italy
\\$^4\,$Department of Astronomy \& Astrophysics, University of California, 1156 High Street, Santa Cruz, CA 95064, USA
\\$^5\,$Institut d'Astrophysique de Paris, Sorbonne Universit{\'e}s, UPMC Univ Paris 6 et CNRS, UMR 7095, 98 bis bd Arago,\\ 75014 Paris, France}
\date{\today} 
\maketitle

\begin{abstract} 
Recent observations suggest that hydrogen reionization ends late ($z \simeq 6$) and proceeds quickly. We present here a new model of the meta-galactic UV/X-ray background (UVB) that is consistent with this. It adopts the most recent determinations of the ionizing emissivity due to stars and AGN, as well as of the \HI absorber column density distribution. Another major improvement is a new treatment of the intergalactic medium (IGM) opacity for ionizing photons that is able to consistently capture the transition from a neutral to an ionized IGM. Previous synthesis models of the UVB, when used in simulations, yield reionization and thermal histories that are inconsistent with the assumed ionizing emissivities. With our new treatment, this discrepancy is fully resolved. In our fiducial model, galaxies leaking $\lta 18\%$ of their Lyman continuum emission drive \HI reionization, while AGN drive \HeII reionization (completing at $z \simeq 6.2$ and $2.8$ respectively). Due to the limited time available for cooling between \HI and \HeII reionization, higher IGM temperatures are predicted for late reionization scenarios. In our fiducial model, the predicted temperatures agree well with observational constraints at $z \lesssim 4$, while being slightly high compared to (somewhat uncertain) data above that. Models with a larger contribution of AGN are instead disfavoured by the temperature data, as well as by measurements of the \HI and \HeII \lya forest opacities. We also present ``equivalent-equilibrium" ionization/heating rates that mimic our fiducial UVB model for use in simulation codes that assume ionization equilibrium.
\end{abstract}


\section{Introduction}
\label{sec:introduction}
The reionization of the all-pervading intergalactic medium (IGM) is a landmark event in the history of structure formation in the Universe, and represents the last global phase transition in the cosmic evolution of baryons. 

In the last decade, several studies of \lya absorption in the spectra of distant quasars showed that hydrogen is fully ionized at least out to   
$z \simeq 5.6$, while inhomogeneous reionization may be still ongoing at slightly earlier times, $z\simeq 6$ \citep[e.g.,][]{fan06b,songaila04,becker15,mcgreer15}. Such a ``late reionization" scenario is supported by recent observations of cosmic microwave background (CMB) anisotropies and polarization, which favour a low Thomson scattering optical depth towards the CMB, $\tau_{\mathrm CMB} = 0.058 \pm 0.012$ \citep{planck16reion}. It is also consistent with the drop in the observed abundance of \lya emitting galaxies at $z \gtrsim 6$ \citep[e.g.,][]{treu13,zheng17}, which is (at least partly) due to absorption by an increasingly neutral IGM \citep[e.g.,][]{choudhury15,mesinger15}. Further evidence for late reionization comes in the form of surprisingly large observed fluctuations in the \lya forest opacity averaged on large scales at $z \gtrsim 5.5$ \citep{becker15}. Although the exact mechanism by which they are generated is still debated, they are most likely a relic of a recently completed patchy reionization. 

While it is understood that the IGM is reionized and kept ionized by the integrated EUV emission from AGN and star-forming galaxies, there is still no 
consensus on the relative contributions of these sources as a function of cosmic epoch.
Because of the high ionization threshold and small photoionization cross section of \hbox{He~$\scriptstyle\rm II$}, and of the rapid recombination rate of 
\hbox{He~$\scriptstyle\rm III$}, the double ionization of helium is expected to be completed by hard UV-emitting quasars around the apparent peak of their activity, at $\simeq 2.5$ \citep[e.g.,][hereinafter HM12]{mcquinn09,HM12}, at much later times than the reionization of \HI and \HeI. Recent data, however, suggest that double helium ionization may actually have occurred earlier than expected \citep{worseck16}, though such conclusion needs to be further confirmed by independent studies. At $z\gta 3$ the apparent declining population of bright quasars makes an increasingly small contribution to \HI ionizing radiation, as it is generally believed that massive stars in relatively faint 
star forming galaxies provide the bulk of the UV background at such early epoch \citep[e.g.,][]{MHR99, meiksin05, faucher08a}.

The galaxy plus AGN picture has been recently questioned by \citet{MH15} (hereinafter MH15), who showed how a significant population of previously undetected faint AGN at $4\lta z\lta 6.5$, claimed by \citet{giallongo15}, may actually dominate \HI reionization 
\citep[albeit with a somewhat extreme choice of parameters; see, e.g.,][]{FHsalvaterra15}. 
Normal star-forming galaxies would then provide, in this scenario, only a negligible contribution to the UVB at all epochs. Interestingly at the same time \citet{chardin2015,chardin2017} argued that such a significant contribution of AGN to the ionizing emissivity at high redshift may explain the observed large \lya forest opacity fluctuations on large scales \citep{becker15}.

The thermal history of the IGM, which reflects the timing and duration of the H and He reionization processes, as well as the nature of the ionizing sources (``hard" QSO-like vs. "soft" stellar-like spectra), is a key tool for understanding the epoch of reionization (EoR). Measurements of the IGM temperature in the range $2\lta z \lta 6$ show a non-monotonic trend, calling for a substantial injection of energy \citep{becker11, boera14, bolton12, bolton14,upton16}. 
Such energy injection is generally ascribed to \HeII photoionization \citep{bryan00,ricotti00,schaye00,mcdonald01,zaldarriaga01,lidz10,becker11,bolton12,garzilli12}, though different sources may play a role \citep[see, e.g.,][]{puchwein12, madaufragos17}. 

Once this vast body of observational evidences is considered, a mild 
tension between the late reionization scenario hinted by CMB and \lya emitter observations, and the evolution of the IGM 
temperature with redshift, does emerge. Therefore, any physically motivated UV background (UVB) model is necessary tightly constrained. 

Spatially homogeneous UVB models are widely used in cosmological hydrodynamical simulations of cosmic structure formation as a simple and efficient way of following the photoionization and photoheating of the IGM during cosmic reionization. As shown in \citet{puchwein15} and explicitly investigated in \citet{onorbe17}, such models, however, result in reionization histories in simulations that are inconsistent with the ionizing emissivities assumed in the UVB models. This typically results in a too early reionization and heating of the IGM, e.g., for the HM12 model most of the temperature increase happens at $z\gtrsim14$. Such artificial early heating will also affect high-redshift star formation in low-mass galaxies.  

We propose here a new fiducial UVB model, that resolves this issue and is able to account for the many existing observational constraints.
To do so, we improve our 1-D radiative transfer code CUBA \citep[][HM12]{HM96}, upgrading the source function with the most recent 
determinations of star formation rate and AGN evolution across cosmic time. We further update the sink term with new  
observational constraints of the mean free path of Lyman continuum photons at $z\gta 3$, at the same time employing a different 
recipe for dust corrections to the observed UV emissivities. The major improvement that results in consistent reionization histories is, however, a novel scheme for treating the \HI and \HeII attenuation along the line of sight, which allows us to asses the mean UVB {\it before and after} the reionization of H and He. 

The paper is organized as follows: We describe our methods of modelling the UVB and the thermal and ionization evolution of the IGM in Sec.~\ref{sec:methods}. In Sec.~\ref{sec:new_model}, we present our new UVB model and the evolution of IGM properties that it implies. We then discuss models with a significant AGN contribution to \HI reionization and test them against data in Sec.~\ref{sec:agn_models}. Finally, a summary of our results is provided in Sec.~\ref{sec:conclusions}.  


\section{Methodology}
\label{sec:methods}
\subsection{Cosmological radiative transfer}
\label{sec:cosmoRT}
We start by briefly reviewing the basic equations governing the propagation of ionizing radiation in a  clumpy medium.
For a thorough discussion we refer to \citet{HM96}, \citet{MH09} and HM12.  

The equation 
of cosmological radiative transfer (RT) describing the time evolution of the 
space- and angle-averaged specific intensity $J_\nu$ is
\begin{equation}
\left({\partial \over \partial t}-\nu H {\partial \over \partial \nu}\right)J_\nu+3HJ_\nu=
- c\kappa_\nu J_\nu + {c\over 4\pi}\epsilon_\nu, 
\label{eq:rad}
\end{equation}
where $H(z)$ is the Hubble parameter, $c$ the speed of the light, $\kappa_\nu$ is the 
absorption coefficient, and $\epsilon_\nu$ the specific proper volume emissivity. The integration of 
Eq.~(\ref{eq:rad}) gives the background intensity at the observed frequency
$\nu_o$, as seen by an observer at redshift $z_o$, 
\begin{equation}
J_{\nu_o}(z_o)={c\over 4\pi}\int_{z_o}^{\infty}\, |dt/dz| dz
{(1+z_o)^3 \over (1+z)^3} \epsilon_\nu(z) e^{-\bar\tau},
\label{eq:Jnu}
\end{equation}
where $\nu=\nu_o(1+z)/(1+z_o)$, $|dt/dz|=H^{-1}(1+z)^{-1}$, and $\bar\tau\equiv -\ln \langle e^{-\tau}\rangle$ 
is the effective absorption optical depth of a clumpy IGM. In the case of Lyman continuum absorption by Poisson-distributed systems, the effective opacity between $z_o$ and $z$ is
\begin{equation}
\bar\tau(\nu_o,z_o,z)=\int_{z_o}^z\,
dz'\int_0^{\infty}\, dN_\nHI\, f(N_\nHI,z') (1-e^{-\tau_c}), 
\label{eq:taueff}
\end{equation}
where $f(N_\nHI,z')$ is the bivariate distribution of absorbers in redshift and
\HI column density along the line of sight, $\tau_c$ is the continuum optical depth at frequency
$\nu'=\nu_o(1+z')/(1+z_o)$ through an individual absorber,
\begin{equation}
\tau_c(\nu')=N_\nHI\sigma_\nHI(\nu') +N_\nHeI\sigma_\nHeI(\nu') +N_\nHeII\sigma_\nHeII(\nu'),
\label{eq:tauC}
\end{equation}
where $N_i$ and $\sigma_i$ are the column densities and photoionization cross sections of ion $i$.

Besides photoelectric absorption, resonant absorption by the hydrogen and helium  Lyman series will produce 
a sawtooth modulation of the radiation spectrum \citep{haiman97,MH09}. Absorption and recombination 
re-emission from the Lyman series of \HI and \HeII can be accounted for as detailed in \citet{MH09} and HM12.

\subsection{The IGM effective opacity across reionization}
\label{sec:uvb_modelling}
A modification of the formalism just described is needed when considering the consistency of the solution of the RT equation to the constraints imposed by the photon budget required to ionize the IGM. In \citet{HM96} and HM12 our original aim was at an empirical assessment of the UVB based on observed quantities, such as the luminosity functions of ionizing sources and the \HI column density distribution of discrete absorbers, $f(N_\nHI,z)$. In particular, in solving Eq.~(\ref{eq:Jnu}) this latter was not derived through photoionization calculations from an underlying H distribution, rather it is an input quantity, alongside the source volume emissivity. Indeed, in HM12 we forced the \HI effective opacity to dramatically increase above a pre-assigned ``reionization" redshift ($\simeq 6$), with the specific aim of mimicking the sharp increase of the \lya effective opacity observed in the IGM at this epoch \citep[see, e.g.,][]{fan06b}. 

In recent years, however, numerical studies \citep{puchwein15,onorbe17} have shown that the pre-assigned emissivity and opacity employed in HM12 do combine in a model UVB which in turn produces an ionization history of the IGM in cosmological simulations that is not consistent with the assumed ionizing emissivity. The origin of this discrepancy is that the neutral hydrogen abundance corresponding to the input \HI opacity, when extrapolated to very high-$z$ (e.g., to the beginning of reionization), is inconsistent with the mean hydrogen density of the Universe.

A further relevant aspect to consider involves the implicit assumptions underlying the RT formalism itself. As a matter of fact, Eq.~(\ref{eq:Jnu}) implicitly assumes that the source function 
(i.e., $\epsilon_\nu$) is meaningfully defined on scales as short as the mean free path of ionizing photons. Such approximation clearly breaks down when the mean free path is comparable, or shorter, than the average distance between ionizing sources. It is therefore not obvious that the solution of the RT equation itself could correctly reproduce the sudden increase of the photon mean free path expected when \HII regions overlap during reionization. This point was raised by, e.g., \citet[][]{davies14}, who showed that the discrete nature of the sources gives rise to a radiation field whose volume-averaged intensity is actually lower than the value computed by solving the RT equation. The basic reason is that, in the pre-reionization Universe, the mean free path is dominated by the still neutral IGM embedding detached \HII regions. Such a completely neutral component does not exist in the case of a continuously distributed source function. 

The considerations above led us to consider in Eq.~(\ref{eq:Jnu}) a modified sink term, tailored in such a way that at low $z$ the IGM opacity matches the observational constraints imposed by  
the \lya forest, while at the same time the high-$z$ attenuation approaches that expected in a uniform, neutral medium at mean cosmic density. 
We further constrain the opacity to be consistent with the ionization history of \HI implied by the emissivity employed. 
These two requirements have been implemented in CUBA as detailed in the following. We note that a similar approach has been recently proposed by 
\citet{onorbe17} and more explicitly by \citet[][]{madaufragos17}, based on an earlier work by \citet[][]{mesinger13}. 

We first compute the evolution of the volume filling factor of ionized bubbles $Q$ 
consistently with the assumed ionized emissivity through the reionization equation \citep[for details see, e.g.,][HM12]{MHR99} 
\begin{equation}
\frac{dQ}{dt}=\frac{\dot n_{\rm ion}} {\bar n} -\frac{Q}
{\bar t_{\rm rec}}.
\label{eq:qdot}
\end{equation}
The above reionization equation holds separately for the \HII and \HeIII filling factors. Accordingly, $\bar n$  
indicates the mean H or He number density, while the terms $\bar t_{\rm rec}$ and $\dot n_{\rm ion}$ denote
the volume-averaged recombination timescale and the ionizing photon emissivity relevant to the considered species, respectively. 

A modified version of this equation that explicitly accounts for the presence of optically thick Lyman limit systems and links the pre-overlap with the post-overlap phases of reionization has been recently derived by \citet{madau17}. Below we will use $Q$ only to describe the volume fraction of ionized bubbles, while residual neutral gas in them will be treated separately. We can therefore stick here to Eq.~(\ref{eq:qdot}).     

Note that Eq.~(\ref{eq:qdot}) can be solved before any RT calculation. Indeed $Q$ ultimately depends only upon 
$\epsilon_\nu$ through the ionizing photon emissivity $\dot n_{\rm ion}$ 
(and upon the clumping factor of the ionized IGM, that we take as in HM12). We checked that for plausible emissivities, \HeII reionization lags \HI reionization, i.e., $Q_\nHeIII \leq Q_\nHII$. We further assume that \HI and \HeI reionizations proceed together (i.e., $Q_\nHeII=Q_\nHII$). This allows us to divide 
the IGM in three different environments: a fully ionized IGM composed of \HII and \HeIII; regions where the IGM is in the form of \HeII and \HII; and the neutral IGM 
where only \HI and \HeI exist. 

Within the ionized IGM we consider a residual \HI effective opacity given by the {\it observed} column density distribution $f(N_\nHI,z)$. In each single absorber of given column $N_\nHI$, the resulting $N_\nHeI$ and $N_\nHeII$ are consistently 
computed assuming ionization equilibrium.\footnote{Note that, while the \HI column density distribution is an input quantity of CUBA, the corresponding \HeI and \HeII distributions are not, as they depend upon the computed background intensity $J_\nu$, hence effectively implying an iterative scheme for the solution of Eq.~(\ref{eq:Jnu}).} Following the scheme adopted in HM12, we parametrize the observed \HI column distribution as a piecewise power-law,  
\begin{equation}
f(N_\nHI,z)=A\,N_\nHI^{-\beta}(1+z)^{\gamma}, 
\label{eq:ladis}
\end{equation}
where, compared to HM12, we slightly modify some of the fitting parameters in order to better match new observational constraints \citep[][see Table~\ref{table:fofn} and compare to table 1 of HM12]{prochaska14}. 
At high redshift the distribution is modelled as a simple extrapolation of the low-redshift evolution, i.e., different from HM12 we do not force 
the \HI column density distribution to mimic a sudden increase in opacity above the \HI reionization redshift. 
Note however that in our new formalism the column density distribution is basically irrelevant for small values of $Q$.
\begin{table*}
\begin{tabular}{llllll}
\hline\hline
{Absorbers class} & {log $(N_\nHI/$cm$^{-2})$} & {$\beta$} & {$A$ [cm$^{-2(\beta-1)}$]} & {$\gamma$} & {redshift} \\[3pt]
\hline
\Lya\ forest & $11-16$ & $1.6$ & $10^{8.497}$ & 2.85 & $z>1.56$ \\
& $11-16$ & $1.6$ & $10^{9.595}$ & 0.16 & $z<1.56$ \\
& $16-18$ & $1.95$ & $10^{14.097}$ & $2.85$ & $z>1.56$ \\
& $16-18$ & $1.95$ & $10^{15.195}$ & $0.16$ & $z<1.56$ \\
\hline
\\[-6pt]
LLSs & $18-19.5$ & & & & \\[3pt]
\hline
\\[-6pt]
SLLSs & $19.5-20.3$ & $1.05$ & $10^{-0.456}$ & 1.27 & $z>1.56$ \\
& $19.5-20.3$ & $1.05$ & $10^{-0.003}$ & 0.16 & $z<1.56$ \\[3pt]
\hline
\\[-6pt]
DLAs & $20.3-21.6$ & $2.0$ & $10^{18.829}$ & 1.27 & $z>1.56$ \\
& $20.3-21.55$ & $2.0$ & $10^{19.282}$ & 0.16 & $z<1.56$\\[3pt]
\hline\hline
\end{tabular}
\caption{Parameters of the distribution of intergalactic absorbers.}
\label{table:fofn}
\end{table*}

With the above prescriptions, the modified opacity is then:
\begin{equation}
\begin{split}
\bar\tau(\nu_0, z_0,z)=  & \int_{z_0}^z  dz'\,\left[ Q_\nHeIII(z')\frac{d\bar\tau}{dz'}\Big\rvert_1 \right. \\
+ & [Q_\nHII(z')-Q_\nHeIII(z')]\frac{d\bar\tau}{dz'}\Big\rvert_2 \\
+ &\left. \left[1-Q_\nHII(z')\right]\frac{d\bar\tau}{dz'}\Big\rvert_3 \right].
\label{eq:tautot}
\end{split}
\end{equation}

For the fully ionized IGM (``zone 1") we have the usual effective opacity (compare Eq.~(\ref{eq:taueff})):
\begin{equation}
\frac{d\bar\tau}{dz'}\Big\rvert_1=\int_0^\infty{d N_\nHI \, f(N_\nHI,z')(1-\rm{e}^{-\tau_c})},
\label{eq:tau1}
\end{equation}
where the continuum optical depth through an individual cloud $\tau_c$ is given by Eq.~(\ref{eq:tauC}). 

In regions where H is ionized and He is single ionized (``zone 2"), the opacity is given by 
\begin{equation}
\frac{d\bar\tau}{dz'}\Big\rvert_2= \bar n_\nHe\, \sigma_{\nHeII}(\nu')\, \frac{d\ell}{dz'} + \int_0^\infty{d N_\nHI \, f(N_\nHI,z')(1-\rm{e}^{-\tau_c})},
\label{eq:tau2}
\end{equation}
where $d\ell/dz'$ is the line element in a Friedmann cosmology, and $\bar n_\nHe$ is the mean cosmic He density. 
In the considered environment, the $N_\nHeI$ and $N_\nHeII$ columns of a cloud of a given $N_\nHI$ should be computed adopting an ionizing radiation field whose flux above 4 Ry is set equal to zero. However, we note that the above Eq.~(\ref{eq:tau2}) produces an excess in He absorption, as it is 
essentially ``double counted" (in the discrete systems {\it and} in the diffuse IGM). In order to avoid this issue, we simply set the He content of the discrete clouds lying in zone 2 equal to zero. 

Finally, in the fully neutral IGM (``zone 3") 
\begin{equation}
\frac{d\bar\tau}{dz'}\Big\rvert_3= \bar n_\nH\,\sigma_{\nHI}(\nu')\,\frac{d\ell}{dz'} \,+\, \bar n_\nHe\, \sigma_{\nHeI}(\nu')\,\frac{d\ell}{dz'}, 
\label{eq:tau3}
\end{equation}
where $\bar n_\nH$ is the mean cosmic hydrogen density.

\begin{figure}
\centerline{\includegraphics[width=\linewidth]{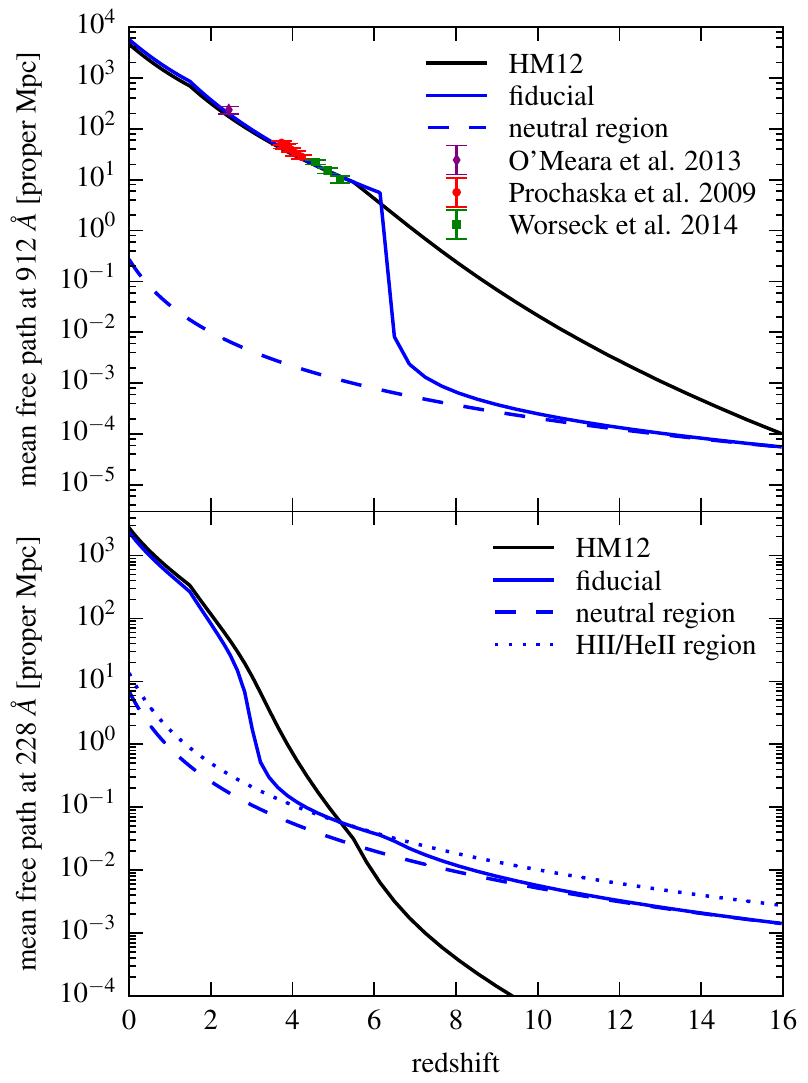}}
\caption{Proper mean free path for photons with wave lengths of 912 \AA (top panel) and 228 \AA (bottom panel) as a function of redshift. Results are shown for our new fiducial model and the HM12 model. For comparison the mean free path expected in a neutral IGM at mean cosmic density is indicated. In the lower panel, the mean free path in a mean-density region containing only \HII and \HeII is also shown. In the upper panel observational constraints from \citet{prochaska09b}, \citet{omeara13} and \citet{worseck14} are shown for reference.}
\label{fig:mfp}
\end{figure}

The mean free path for ionizing photons at 912 and 228~\AA\ implied by our new opacity prescription is shown in Fig.~\ref{fig:mfp} as a function of redshift and compared to the HM12 model. With our new prescription the mean free path at 912 and 228 \AA\ drops much more strongly when entering the epochs of \HI and \HeII reionization, respectively. The expected values in a neutral region at mean density are also indicated for comparison. 

\subsection{The sources driving reionization}
\label{sec:sources}
The only sources of ionizing radiation included in CUBA are star-forming galaxies and AGN, as it is generally accepted that they are the main drivers of 
the reionization process in the IGM. We model these two populations separately, along 
the lines detailed in HM12 (we refer to that paper for full details), with the modifications we describe in the following. 

\subsubsection{Ionizing emission from galaxies}

The comoving, dust-reddened frequency-dependent emissivity as a function of cosmic time $t$ arising from the population of star forming galaxies is computed as a 
convolution integral:
\begin{equation}
\epsilon_{\nu,c}(t)=C(t)\int_0^t{d\tau\, {\rm SFRD}(t-\tau)\, l_\nu[\tau, Z(t-\tau)]}.
\label{eq:eps_gal}
\end{equation}
We use a modified version of the best-fitting comoving star formation rate density (SFRD) proposed by \citet{madaufragos17}. For a Kroupa IMF \citep{kroupa01} 
our best-fit model reads
\begin{equation}
{\rm SFRD}(z)=0.01\frac{(1+z)^{2.7}}{1+[(1+z)/3.0]^{5.35}}\,\sfrd.
\label{eq:SFRD}
\end{equation}
The above function is an updated version of eq. (15) of \citet{madaudickinson14}, and 
reproduces recent results in the range $4\lta z \lta 10$ \citep{bowler15, finkelstein15, ishigaki15, mcleod15, oesch15, mcleod16}. The above 
SFRD is consistent in the redshift range $5.5\lta z\lta 10$ with the UV luminosity densities derived by \citet{bouwens15} under the assumption of a faint-end luminosity function cut-off of $M_{\rm lim}=-13$ mag, deeper than what was assumed by \citet{madaufragos17} (i.e., $M_{\rm lim}=-16$). Eq.~(\ref{eq:SFRD}) has to be compared with the corresponding eq. (53) in HM12.  

In Eq.~(\ref{eq:eps_gal}), the function $l_\nu[\tau, Z(t-\tau)]$ is the 
specific luminosity per unit initial stellar mass radiated by a single stellar population (SSP) at age $\tau$ and metallicity $Z(t-\tau)$, that we compute using the code STARBURST99 \citep{leitherer99, vazquez05, leitherer10, leitherer14}. We use SSPs of decreasing metallicities with redshift according to 
\begin{equation}
\log\langle Z/Z_\odot \rangle =0.153 - 0.074 z^{1.34},      
\label{eq:Zvsz}
\end{equation}
to be compared to eq. (52) in HM12. Eq.~(\ref{eq:Zvsz}), proposed by \citet{madaufragos17}, represents the best fit to the data of the mass-weighted metallicity of newly formed stars, obtained by integrating the mass-metallicity relation of \citet{zahid14} over the evolving galaxy stellar mass function of \citet{baldry12}, \citet{ilbert13}, \citet{kajisawa09}, \citet{lee12} and \citet{grazian15} in the range $0\lta z \lta 7$. 

Finally, the time-dependent pre-factor $C(t)$ in Eq.~(\ref{eq:eps_gal}) represents the escape fraction of Lyman continuum photons for energies above 1 Ry, 
while it accounts for dust absorption below. The escape fraction $f_{\rm esc}$ is the single, most crucial parameter in the modelling of the 
ionizing emissivity from galaxies. 
In our fiducial model, we adopt the same functional form as in HM12, but with different numerical parameters:
\begin{equation}
  f_{\rm esc} = \min \left[6.9\times 10^{-5} \, (1+z)^{3.97}, \, 0.18 \right],
\label{eq:fesc}
\end{equation}
while dust attenuation is treated using a \citet{calzetti00} attenuation law, where we adopt the following fit to the luminosity 
weighted attenuation at 1500 \AA\ :
\begin{equation}
A_{\rm FUV}(z)=2.5 \log \left[\frac{3.5+5.5(z/0.85)^{1.5}}{1+(z/1.45)^{3.0})}\right].
\label{eq:AFUV}
\end{equation}
This expression (to be compared to eq. (50) in HM12) fits the compilation of data in the range $0\lta z\lta 8$ presented by \citet[][see their table 1 and references therein]{madaudickinson14}. 
Note that, since we correct the SSP emissivity for dust 
absorption, $f_{\rm esc}$ defined in Eq.~(\ref{eq:fesc}) has to be considered an {\it absolute escape fraction}, rather than the relative one usually derived from observations \citep[see, e.g.,][]{rutkowski16}.

\begin{figure}
\centerline{\includegraphics[width=\linewidth]{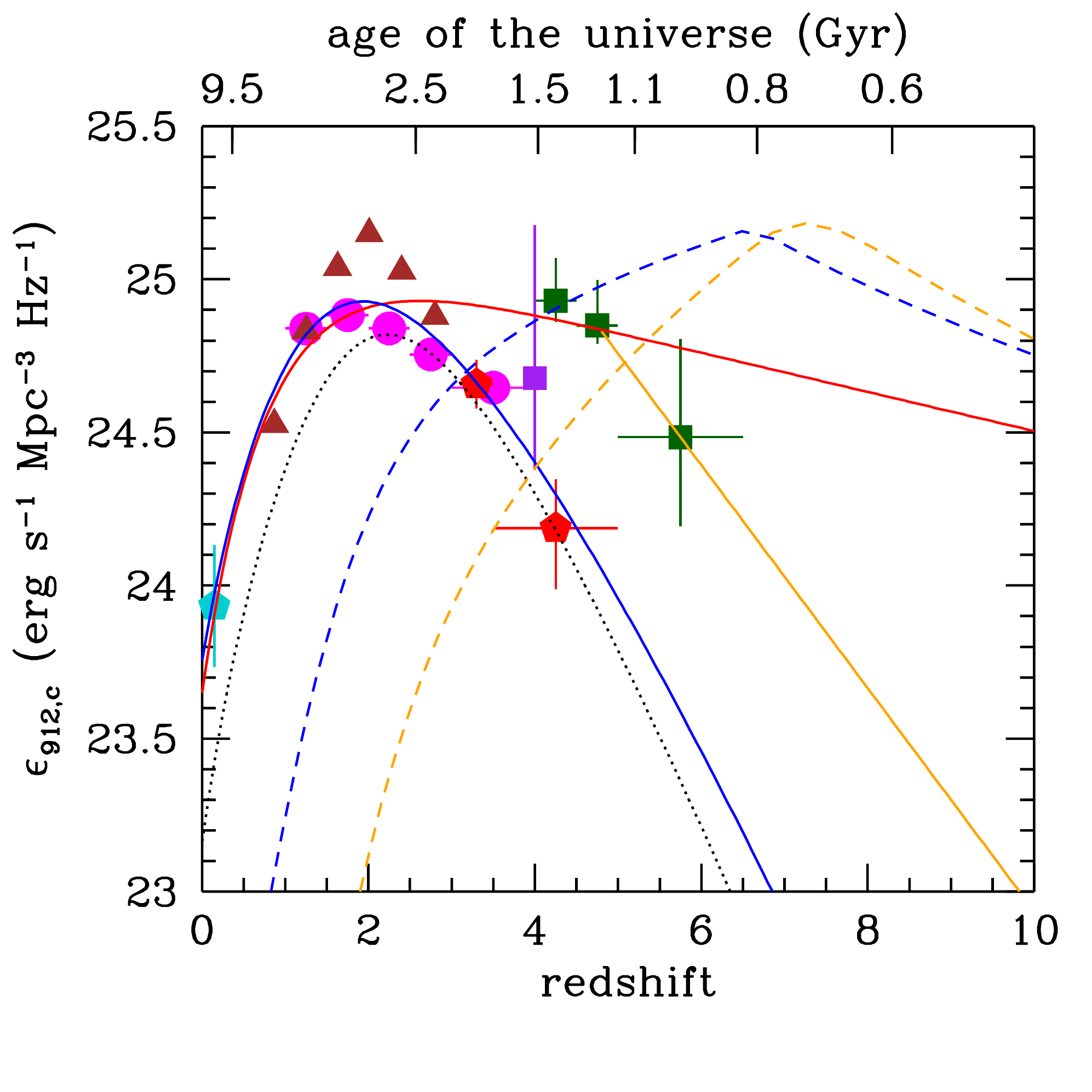}}
\caption{Comoving emissivity at 912 \AA \, as a function of redshift. Data points refer to AGN contributions as inferred by \citet[][cyan pentagon]{schulze09}, 
\citet[][brown/dark red triangles]{palanque13}, 
\citet[][magenta circles]{bongiorno07}, \citet[][red pentagons]{masters12}, \citet[][purple square]{glikman11}, and \citet[][green squares]{giallongo15}. For comparison, 
the Lyman continuum emissivity from AGN of \citet{hopkins07} is shown as a black dotted line. The solid blue line is the AGN emissivity given by Eq.~(\ref{eq:mod_Hopk}) that, 
combined with the emissivity from galaxies (shown as a blue dashed line), defines our fiducial UVB model. The emissivity proposed by MH15 for a 
UVB dominated by AGN is shown as a red solid line. We also show a case where the AGN contribution is taken from the \citet{giallongo15} data at face value (orange solid line at $z\geq 4.75$), with a further contribution at high $z$ from star forming galaxies (orange dashed line, see Sec.~\ref{sec:agn_models} for details).}
\label{fig:eps912}
\end{figure}

\subsubsection{Ionizing emission from AGN}
AGN are the other source of ionizing radiation considered. The comoving emissivity at 912 \AA, $\epsilon_{912,c}$ (in units $\lumdens$), is modelled as 
\begin{equation}
\epsilon_{912,c}(z)= 10^{24.6}\,(1+z)^{4.2}
\,{\exp(-0.3z)\over \exp(1.5z)+6.0}. 
\label{eq:mod_Hopk}
\end{equation}
Such functional form, shown in Fig.~\ref{fig:eps912} as a solid blue line, fits reasonably well the emissivity at $z\simeq 0$ from \citet{schulze09}, at $1\lta z \lta 3.5$ 
from  \citet{bongiorno07}, and at $3.5 \lta z \lta 4.5$ from \citet{masters12}. The 1 Ry emissivity data were inferred by MH15 
by converting the optical emissivity derived in the mentioned studies, using a power-law spectral index, $\alpha_{\rm UV}=-0.61$, after \citet{lusso15}. 
For $z\lta 2$, Eq.~(\ref{eq:mod_Hopk}) closely follows the emissivity proposed by MH15 for a UVB dominated by AGN, while for $z\gta 2$ it lies just above the emissivity used in HM12 following \citet{hopkins07} (see eq. (37) in HM12). 

The UV spectral energy distribution at wavelengths shorter than 912 \AA\ is modelled as a power law, 
$f_\nu \propto d L / d \nu \propto \nu^{\alpha_{\rm UV}}$, with $\alpha_{\rm UV}=-1.7$. Finally, we update the HM12 prescription for the X-ray properties of AGN adopting the luminosity dependent density evolution of the [2-10]keV luminosity function  proposed by \citet{ueda14}.

\setlength{\tabcolsep}{4pt}
\begin{table*}
\centering
\resizebox{\linewidth}{!}{\hspace{-6mm}
\begin{tabular}{lccccccccccc}
\hline\hline
model & AGN        & AGN spectral & SFRD & $f_{\rm esc,0}$  & $\alpha_{\rm esc}$ & $f_{\rm esc,max}$ & $z_{\rm HI}$ & $f_{\rm esc,HeII}$ & $z_{\rm HeII}$ & new & $\tau_{\rm CMB}$ \\
name  & emissivity & slope & & & & & & & & eff. $\tau$ & \\
\hline\hline
 HM12 & eq.~(37) in HM12 & -1.57 & eq.~(53) in HM12 & $1.8\times10^{-4}$ & 3.4 & 1.0 & 10.3 & 1.0 & 3.3 & N & 0.122 \\
\hline
 AGN-dominated & eq.~(1) in MH15 & -1.7 & - & - & - & - & 5.6 & 1.0 & 3.9 & Y & 0.067 \\
\hline
 AGN-assisted & see caption & -1.7 & eq.~(1) in MF17 & $1.7\times10^{-7}$ & 7.2 & 0.5 & 6.8 & 1.0 & 3.7 & Y & 0.068 \\
 AGN-assisted, $f_\mathrm{esc,HeII}=0.5$ & see caption & -1.7 & eq.~(1) in MF17 & $1.7\times10^{-7}$ & 7.2 & 0.5 & 6.8 & 0.5 & 2.7 & Y & 0.068 \\
 AGN-assisted, $\alpha_{\rm UV}=-2$ & see caption & -2.0 & eq.~(1) in MF17 & $1.7\times10^{-7}$ & 7.2 & 0.5 & 6.7 & 1.0 & 2.9 & Y & 0.068 \\
\hline
fiducial & Eq.~(\ref{eq:mod_Hopk}) & -1.7 & Eq.~(\ref{eq:SFRD})& $6.9\times10^{-5}$ & 3.97 & 0.18 & 6.2 & 1.0 & 2.8 & Y & 0.065 \\
\hline\hline
\end{tabular}%
}
\caption{Summary of the source and reionization models considered in this work. The columns list the model name, the prescription of the AGN ionizing emissivity at 912 \AA, the AGN spectral slope for Lyman continuum radiation, the star formation rate density, the (z=0) normalization of the escape fraction of ionizing photons emitted by stars, the power-law index for the redshift evolution of the escape fraction, the maximum value of the escape fraction, the redshift at which \HI reionization finishes, the escape fraction of \HeII ionizing photons, the redshift at which \HeII reionization finishes, whether the new effective intergalactic opacity treatment for ionizing photons was used (Y(es)/N(o)), and the Thomson scattering optical depth towards the CMB predicted by the model, respectively. The AGN ionizing emissivity at 912 \AA\ in all AGN-assisted models is the same as in the AGN-dominated model for $z<4.75$, while for $z>4.75$ we simply adopted a straight line through the two higher redshift \citet{giallongo15} data points in Fig.~\ref{fig:eps912}. \citet{madaufragos17} has been abbreviated as MF17 in this table.}
\label{tab:sims}
\end{table*}
\setlength{\tabcolsep}{6pt}
%

\subsection{Modelling the IGM ionization and temperature}
\label{sec:ion_temp_modelling}

The numerical modelling of the IGM ionization state and temperature follows \citet{puchwein15}. Equipped with the photoionization and photoheating rates
from the cosmological radiative transfer calculation, we integrate the ionization and recombination rate equations for hydrogen and helium (accounting for photoionization, recombination, and collisional ionization) as well as the heating and cooling rate equations (accounting for photoheating, free-free cooling, collisional excitation and ionization cooling, recombination cooling, Compton cooling and Hubble cooling). We start with a cold and neutral IGM at high redshift and evolve
the IGM temperature and ionization state down to redshift zero. Importantly, unless explicitly stated otherwise, we solve the full non-equilibrium evolution equations, thus allowing for deviations from ionization equilibrium which typically occur during the epochs of hydrogen and helium reionization. A more detailed discussion of such non-equilibrium effects is given in \citet{puchwein15}.

In the cosmological hydrodynamical simulations presented below, we computationally efficiently integrate this set of stiff ordinary differential equations with the CVODE library\footnote{http://computation.llnl.gov/casc/sundials/main.html} \citep{Cohen1996,Hindmarsh2005}, while for predicting the evolution of the IGM properties at mean cosmic density we also use a simpler code that performs a direct explicit integration with a very large number of time steps, $\sim 10^6$. The agreement of the predictions with full cosmological simulations is demonstrated in Appendix~\ref{sec:cosmo_vc_one-cell}.  

Nevertheless our modelling of the IGM thermal and ionization state during the epoch of reionization is still strongly simplified. In particular, since we apply the same homogeneous UV background everywhere in the cosmological simulations, we do not properly capture the patchy nature of reionization in which the IGM consists of neutral regions and ionized bubbles. In Appendix~\ref{sec:temp_modelling_patchy} we asses how sensitive our predictions are do this approximation and conclude that they should be reasonably accurate after the end of hydrogen reionization where essentially all of the observational data falls that we compare to.      

\section{The new UV background model}
\label{sec:new_model}

\begin{figure*}
\centerline{\includegraphics[width=\linewidth]{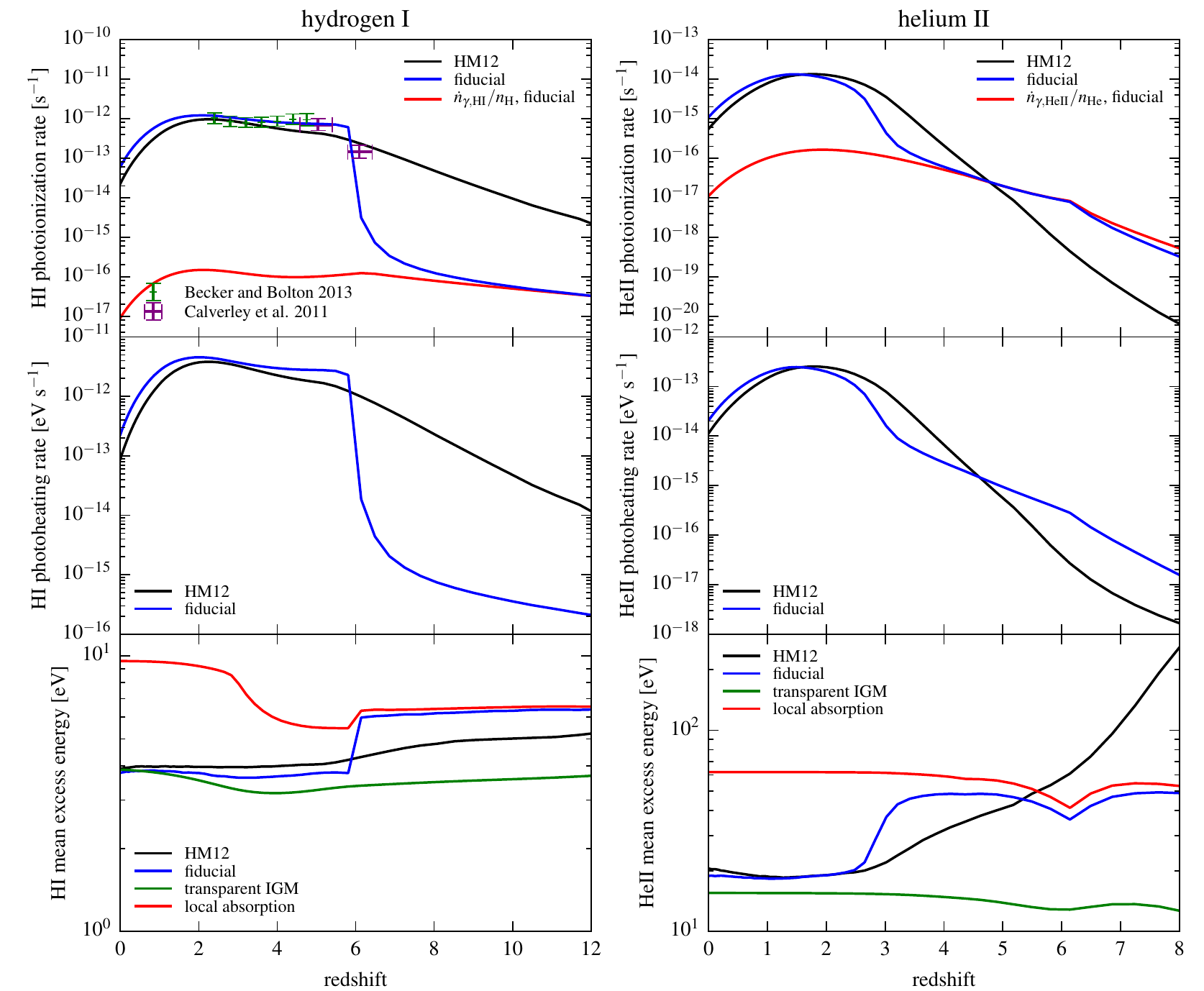}}
\caption{Photoionization (top) and photoheating (middle) rates and mean excess energy (bottom) for \HI (left) and \HeII (right) in our fiducial UVB model. The HM12 model is shown for comparison. At high redshift in the epoch of reionization the photoionization rates drop faster when using the new scheme for the IGM effective opacity. As expected from photon number conservation, the \HI/ \HeII photoionization rates approach the \HI/ \HeII ionizing photon production rates per hydrogen/helium atom (including all ionization states) at high $z$ in the fiducial model (red curves in top panels). In the upper left panel, observational constraints from \citet{calverley11} and \citet{becker13} are shown for comparison. The behaviour of the photoheating rates is qualitatively similar to the photoionization rates. For the fiducial model the mean excess energy (beyond the ionization threshold) is compared to the expected excess energy in the limit of complete local absorption of the ionizing photons, as well as in the limit of a transparent IGM without spectral filtering.}
\label{fig:rates}
\end{figure*}

In this section, we will present the properties of the UVB when modelled as detailed in Sec.~\ref{sec:methods}, i.e. in particular with our new prescription of the effective opacity. We will also assess the implications that our new fiducial UVB model has for the evolution of the thermal and ionization state of the IGM. Some of the basic parameters of this model are summarized in Table~\ref{tab:sims} and compared to the parameters of the HM12 model, as well as of the models that will be presented in Sec.~\ref{sec:agn_models}. 

\subsection{Photoionization and heating rates}
\label{sec:rates}

The different treatment of the effective opacity for ionizing photons will of course affect the UV background in particular for photon energies larger than the ionization threshold and during the reionization of the relevant ion species. This is illustrated in Fig.~\ref{fig:rates}, while the numerical values of the photoionization and photoheating rates are provided in Table~\ref{tab:rates}. The hydrogen photoionization rate, $\Gamma_\mathrm{HI}$, in our new fiducial model is roughly similar to what was found for the HM12 model after the end of hydrogen reionization, i.e. for $z \lesssim 6$. During hydrogen reionization, at $z \gtrsim 6$, the effective opacity for hydrogen ionizing photons is however significantly larger compared to HM12. This also results in much lower values of $\Gamma_\mathrm{HI}$ in the new fiducial model.

It is worth considering the meaning of these lower photoionization rates during reionization for a moment. In reality the photoionization rate during reionization will have a bimodal distribution, being essentially zero in neutral regions, while full radiative transfer simulations suggest that it might hover around $\sim 10^{-13} \, \mathrm{s}^{-1}$ \citep{chardin2015} in ionized bubbles. Clearly, our new rates are not meant to be used as an estimate of the photoionization rate in ionized bubbles during reionization. Instead the rates in our fiducial homogeneous UVB model were chosen to reproduce the number of photoionizations events in an actually inhomogeneously ionized universe. It is, thus, essentially a neutral hydrogen-weighted photoionization rate. To illustrate this, we also show the number of hydrogen ionizing photons that are emitted per second in this model in the upper left panel of Fig.~\ref{fig:rates}. To allow a direct comparison we show this number per hydrogen nucleus, i.e. we divide it by the sum of the numbers of neutral hydrogen atoms and \HII ions (in other words the red curves show the first term, $\dot n_{\rm ion} / \bar{n}$, on the right-hand side of Eq.~(\ref{eq:qdot})). At the beginning of reionization when the mean free path is short and $n_\mathrm{HI} \approx n_\mathrm{H}$, the number of photoionizations events should be given by the number of emitted ionizing photons. Hence, in a homogeneous UVB model $\dot n_{\rm ion} \approx \Gamma_\mathrm{HI} n_\mathrm{HI} \approx  \Gamma_\mathrm{HI} n_\mathrm{H}$ should hold, and thus $\dot n_{\rm ion} / n_\mathrm{H} \approx \Gamma_\mathrm{HI}$. A similar argument can be made for the \HeII photoionization rate and ionizing photon emission rate. The good agreement of the red solid and blue solid curves at high redshift illustrates that the correct limit is indeed recovered in our fiducial UVB model. Note that this was not the case in the HM12 model (compare fig. 1 in \citealt{puchwein15}). The full reionization history implied by the photoionization rates in our fiducial model will be discussed in Sec.~\ref{sec:reionization_history}. 

In the redshift range $0 \leq z \lesssim 6$, the hydrogen photoionization rate is slightly larger than in the HM12 model
and in good agreement with observational constraints. The slightly larger photoionization rate at $z \approx 0$ should help to alleviate tensions between theoretical predictions of the abundance of absorption lines in the local \lya forest and observational constraints \citep[see, e.g.,][]{kollmeier14,viel17, nasir17}. 

Qualitatively the photoionization rates for \HeII during the epoch of \HeII reionization behave similarly. It is
lower in the fiducial model compared to HM12 at $2.7 \lesssim z \lesssim 5$ and also approaches the value expected
from the emission rate of \HeII ionizing photons at $z\sim5$, i.e., near the beginning of \HeII reionization. Somewhat surprisingly $\Gamma_\mathrm{HeII}$ in HM12 falls below our new fiducial model at $z \gtrsim 5$. This can be understood by noting that the mean free path for \HeII ionizing photons in HM12 rather unrealistically drops below what is expected for a universe where all helium is in \HeII as shown in Fig.~\ref{fig:mfp}. 

The middle panels of Fig.~\ref{fig:rates} show the photoheating rates for \HI and \HeII in our fiducial model. For reference the corresponding rates in HM12 are also displayed. Qualitatively, many of the trends seen in the photoionization rates are also present in the photoheating rates. Note that for a fixed UVB spectrum the two would simply be proportional to each other. To understand differences in their evolution better it is instructive to look at the ratio of the photoheating to the photoionization rate, which is just the mean excess energy above the ionization threshold per photoionization event. This energy is available for heating the IGM and shown in the bottom panels of Fig.~\ref{fig:rates}.

Interestingly, the mean excess energy evolves rapidly near the tail end of the reionization of the relevant ion species in the fiducial model, while being largely constant before and after that. In the HM12 model in contrast there is a gradual evolution over a wide redshift range. To understand the evolution of the mean excess energy better it is illustrative to consider two limiting cases: i. complete, local (or immediate) absorption of all ionizing photons and ii. a transparent IGM in which emitted photons simply redshift and no spectral filtering by absorbers occurs. These limits are shown for the fiducial model by the red and green curves in the bottom panels of Fig.~\ref{fig:rates}. Broadly speaking, the mean excess energy in the fiducial model evolves from the complete absorption limit to near the transparent IGM limit at the end of reionization. This behaviour seems plausible given that ionizing photons emitted during the epoch of reionization are expected to be absorbed (at latest) by the first neutral region they encounter, while photons emitted well after reionization will transverse a largely transparent IGM. In comparison, the gradual evolution found in the HM12 model seems somewhat less well motivated. Full details about how the predictions for the two limiting cases were calculated are provided in Appendix~\ref{sec:energy_limits}. It is worth noting here though that there is some uncertainty in the mean excess energy in the complete absorption limit for \HeII, which is related to the maximum photon energy considered. 

\subsection{The spectrum of the UV background}
\label{sec:spec}

\begin{figure}
\centerline{\includegraphics[width=\linewidth]{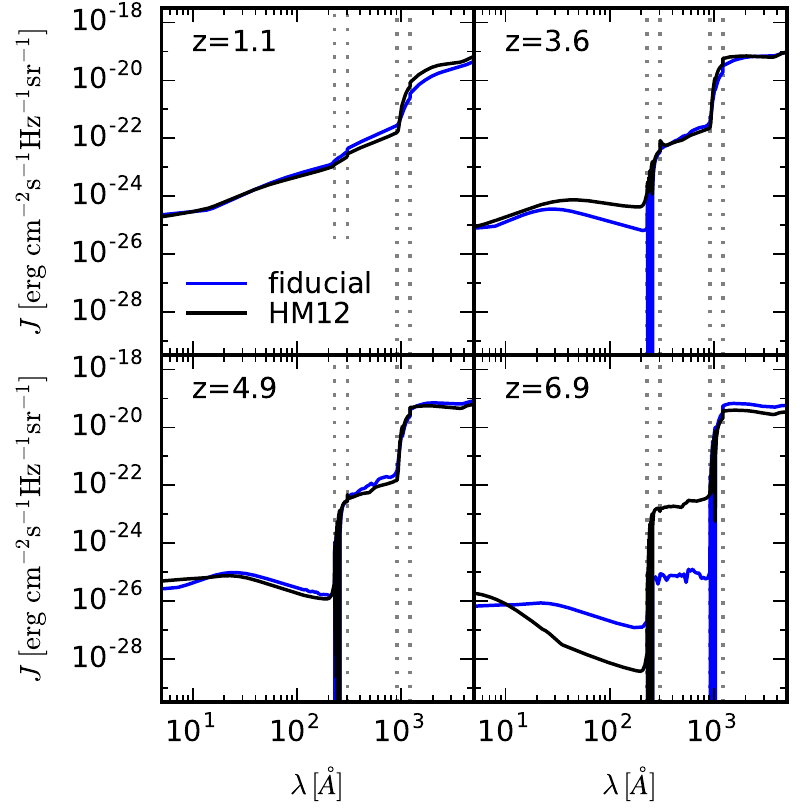}}
\caption{Spectra of the cosmic UV background in our fiducial model and the HM12 model. Results are shown at four different redshifts. 
The vertical dotted lines indicate the \HI and \HeII ionization edges and \lya wavelengths. Between these lines strong sawtooth modulation by the Lyman series of \HI and \HeII is visible in both models before the reionization of the relevant ion species.}
\label{fig:spec}
\end{figure}

Fig.~\ref{fig:spec} displays the cosmic UV background spectrum in our fiducial model at four different redshifts. For reference we also show the HM12 model. Many of the differences discussed in the previous section are also visible here. At redshift 6.9, during the epoch of hydrogen reionization, the fiducial model has a lower intensity at wavelengths shorter than the hydrogen ionization threshold (912 \AA $\,> \lambda \gtrsim$ 300 \AA). This is consistent with the lower hydrogen photoionization rate found at this redshift. The intensity at wavelengths shorter than the \HeII ionization threshold ($\lambda <$  228 \AA) is instead higher, resulting in higher $\Gamma_\mathrm{HeII}$ values. At very short wavelengths though, corresponding to photon energies $\gtrsim 1 \mathrm{keV}$, the intensity in the HM12 model is higher. While this has little effect on $\Gamma_\mathrm{HeII}$, it results in the strong increase of the mean excess energy for \HeII photoionization in the HM12 model at high redshift (compare lower right panel of Fig.~\ref{fig:rates}).

Between \HI and \HeII reionization (see the $z=4.9$ panel), as well as after \HeII reionization ($z=1.1$ panel), the spectrum of the cosmic UVB in our fiducial model is fairly similar to that of the HM12 model. During \HeII reionization ($z=3.6$ panel) the spectra are quite similar for wavelengths longer than the \HeII ionization edge, while the fiducial model has a lower intensity at shorter wavelengths. This is reflected by the lower $\Gamma_\mathrm{HeII}$ values in the fiducial model at these redshifts.
 
\subsection{The reionization history}
\label{sec:reionization_history}

Previous simulation studies which followed photoionization and photoheating with a homogeneous UV background found that reionization happens too early compared to what is expected from the evolution of the ionizing emissivity \citep[e.g.][]{puchwein15,onorbe17}. As discussed above this is caused by an unrealistic continuation of the evolution of the effective opacity of the IGM from lower redshifts into the epoch of reionization. While the opacity is constrained by the observed \HI column density distribution only at $z \lesssim 6$, it needs to be known to $z \sim 15$ to follow cosmic reionization. It is not too surprising that any extrapolation to high redshift is rather uncertain even when a (somewhat ad hoc) increase in the opacity in the epoch of reionization is accounted for as in HM12. To tackle this problem, we have introduced a new method of treating the sinks of ionizing photons before percolation in Sec.~\ref{sec:uvb_modelling}. Here, we investigate how this affects the predicted reionization history.

\begin{figure}
\centerline{\includegraphics[width=\linewidth]{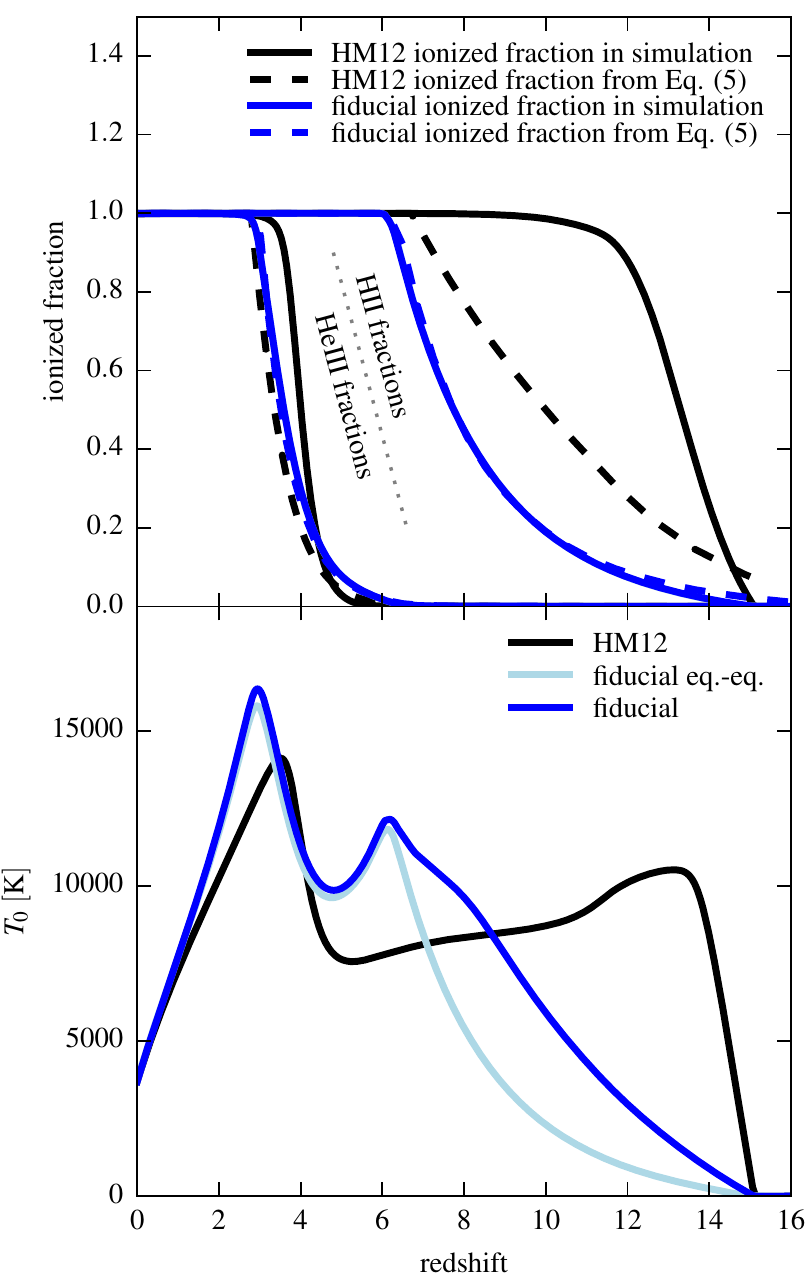}}
\caption{Top panel: \HII (right set of curves) and \HeIII fractions (left set of curves) in simulations (solid) as well as obtained from 
Eq.~(\ref{eq:qdot}) (dashed) are indicated as a function of redshift. Results are shown for the HM12 UV background (black),
as well as for our new fiducial UVB model (blue) which includes the updated treatment of the intergalactic opacity.
In simulations with the latter UVB model, the ionized volume fractions are consistent
with the expectation from Eq.~(\ref{eq:qdot}). Bottom panel: IGM temperature at mean density as a function of redshift. The differences in the reionization history are reflected in the thermal evolution. The heating in our new fiducial model happens later. This results also in a
higher temperature at the tail end of hydrogen reionization, as well as after that. The thermal evolution in a simulation assuming
ionization equilibrium and using the `equivalent-equilibrium' version of our fiducial model (see Sec.~\ref{sec:eq_rates} for further details) is shown in light blue. This model is not visible in the top panel as the curves showing its reionization histories are exactly covered by the blue solid lines.}
\label{fig:Q}
\end{figure}

The top panel of Fig.~\ref{fig:Q} displays the reionization history inferred for our fiducial UVB model as well as for HM12. Shown are the \HII and \HeIII fractions obtained by following photoionization and photoheating with these UVB models. These were obtained with simplified simulations of the thermal evolution of the IGM in which only gas at mean cosmic density is considered. The results are, however, in excellent agreement with full cosmological simulations as demonstrated in Appendix~\ref{sec:cosmo_vc_one-cell}. Importantly, the simulations allow deviations from ionization equilibrium, which is essential for getting an accurate evolution of the thermal and ionization state \citep[see, e.g.,][]{puchwein15}.

We also show the ionization history expected from Eq.~(\ref{eq:qdot}), i.e. from simply considering the ionizing photon budget. For the HM12 model, the simulation results are inconsistent with this expectation. The unrealistic continuation of the evolution of the mean free path from lower redshifts in HM12 results in too large photoionization rates during reionization. This in turn causes a too early reionization when the UVB model is used as a homogeneous background for following photoionization and photoheating in simulations. In contrast, in our fiducial model with the new effective opacity prescription, the simulations and the expectation from the ionizing photon budget are in excellent agreement. 

\subsection{The thermal history of the IGM}
\label{sec:thermal_history}

\begin{figure*}
\centerline{\includegraphics[width=0.8\linewidth]{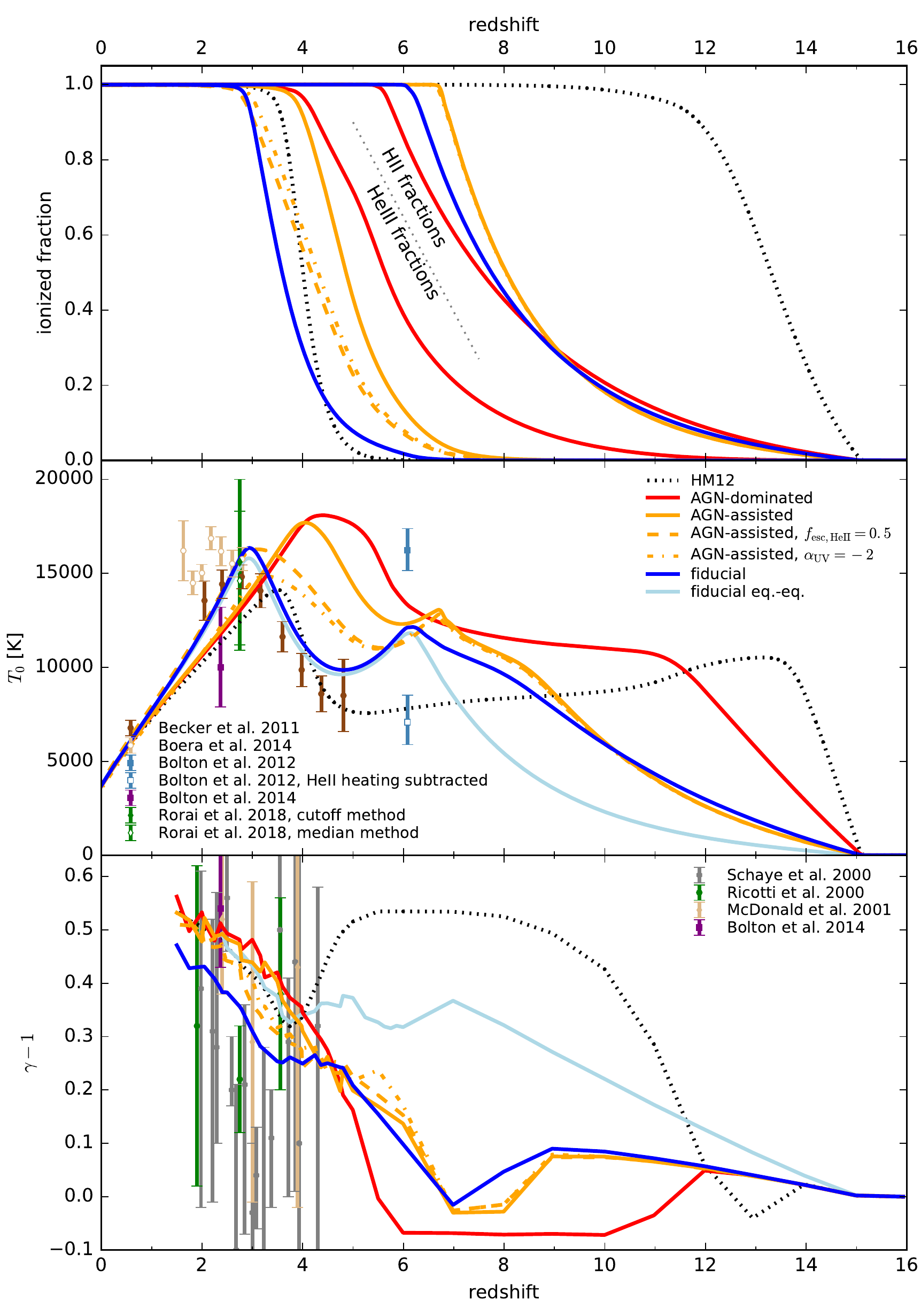}}
\caption{Top panel: \HII fractions (right set of curves) and \HeIII fractions (left set) predicted for our different UVB models as a function of redshift. Middle panel: IGM temperature at mean density. The observational constraints from \citet{becker11} and \citet{boera14} were scaled to mean density using the slope of the density-temperature relation measured from a simulation using the fiducial UVB model. Bottom panel: Slope, $\gamma-1$, of the density-temperature relation. All computations allow for deviations from ionization equilibrium, except the \textit{fiducial eq.-eq.} model which uses the equivalent-equilibrium rates and accordingly assumes ionization equilibrium.}
\label{fig:temp}
\end{figure*}

\begin{figure}
\centerline{\includegraphics[width=\linewidth]{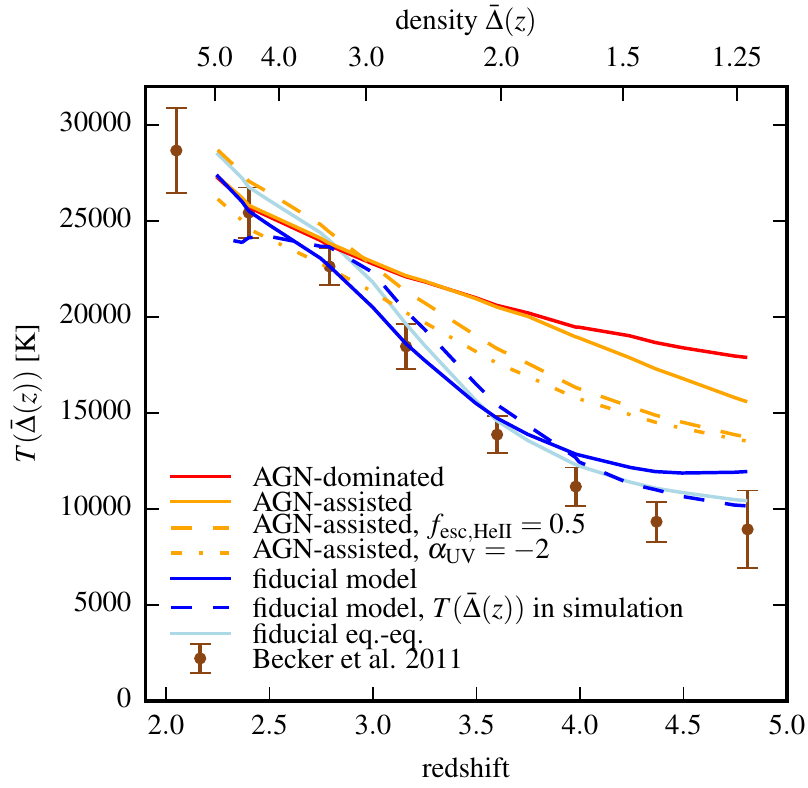}}
\caption{IGM temperature at the overdensity $\bar{\Delta}(z)$ to which the curvature
statistics is most sensitive as a function of redshift. For the simulations with our different UVB models, the temperatures were measured from mock \lya forest spectra using the curvature method introduced in \citet{becker11}. For comparison the observational constraints from that work are also shown. For the fiducial model, we also show the temperature measured at density $\bar{\Delta}(z)$ directly from the gas particles in the simulation (blue dashed) to give an impression of the accuracy of the curvature temperature method.}
\label{fig:temp_Delta}
\end{figure}

The later reionization of hydrogen in our new fiducial UVB model compared to HM12 is reflected in the thermal evolution. The heating from hydrogen reionization happens later. Consequently, there is less time for cooling before \HeII reionization commences and higher temperatures are reached. This is illustrated in the bottom panel of Fig.~\ref{fig:Q} which shows the thermal evolution of the IGM at mean cosmic baryon density as a function of redshift. The temperature in the new fiducial model peaks as expected at the end of hydrogen reionization at $z \sim 6.2$, rather than at $z \sim 13$ as in HM12. In our new model, there is, thus, only a limited period in which the IGM can cool by Hubble expansion and inverse Compton scattering (from $\sim1.2\times10^4$ K to $\sim10^4$ K) before the onset of \HeII reionization at $z \sim 4.5$. This combined with the slightly higher excess energy for \HeII photoionization in the range $2.7 \lesssim z \lesssim 5$ (see Fig.~\ref{fig:rates}) results in higher IGM temperatures all the way down to $z \sim 2$. 

Figs.~\ref{fig:temp} and \ref{fig:temp_Delta} confront the predictions of the thermal state of the IGM for simulations with our new fiducial UVB model (blue solid curves) with observational constraints. Also shown are simulations for UVB models with a larger AGN contribution which will be discussed in Sec.~\ref{sec:agn_models}. The middle panel of Fig.~\ref{fig:temp} compares the predicted IGM temperature at mean density to measurements based on the curvature of \lya forest spectra in the redshift range $1.6 < z < 4.8$ \citep{becker11,boera14}, the lower cutoff of the line width-column density distribution at $z\sim2.4$ \citep{bolton14} and $z\sim2.75$ \citep{rorai18}, the median of the line width-column density distribution at $z\sim2.75$ \citep{rorai18}, and the cumulative line width distribution function in quasar near zones at $z\sim 6$ \citep{bolton12}. The curvature-based measurements are most sensitive at a redshift-dependent density $\bar{\Delta}(z)$. They have been scaled to mean density using the slope of the temperature-density relation from a cosmological hydrodynamical simulation of the fiducial model. The measurements in quasar near zones likely have a contribution from the initial stages of \HeII reionization starting early around these quasars. Both the direct measurements and the temperatures constraints after subtracting the likely contribution from \HeII heating are indicated. Overall, the new fiducial model matches the observational constraints well. An exceptions are the $z \sim 6$ data points based on the quasar near zones, where the predictions fall between the direct measurement and the \hbox{He~$\scriptstyle\rm II$}-corrected constraint. Given the uncertainty in the assumed \HeII heating correction, this is maybe not too worrying.

This overall good agreement with a wide range of IGM temperature data is reassuring. It is nevertheless worth noting that measuring the IGM temperature from the \lya forest is still very challenging. Measurements could potentially be affected by various systematic biases, e.g., due to residual metal line contamination, cosmic variance or a degeneracy between pressure smoothing and thermal broadening of lines. In particular, some IGM temperature measurements are formally inconsistent with each other. For example, the not yet peer-reviewed IGM temperature constraints of \citet{hiss17} are significantly higher than those of \citet{becker11} at $z \gtrsim 2.5$. Clearly, more work is needed to understand the origin of these discrepancies.

The bottom panel of Fig.~\ref{fig:temp} compares the slope $\gamma - 1$ of the temperature density relation, $T \approx T_0 \Delta^{\gamma-1}$, in full cosmological simulations to observational constraints. The slope is measured from the simulations as in \citet{puchwein15}. In our new fiducial model the temperature density relation is almost isothermal down to $z \sim 6$, as the heating from hydrogen reionization is largely independent of density when non-equilibrium effects are taken into account. In contrast in the HM12 model, an isothermal temperature-density relation is only found for $z \gtrsim 12$ due to the pre-mature reionization predicted by that model. Unfortunately, the data on the slope of the temperature-density relation are not very constraining as they have large errors bars and do not extend to high redshift. In reality, the patchy nature of reionization will further complicate the situation. While hot gas at low density is certainly expected, there will be significant spatial fluctuations in the temperature depending on the redshift at which a particular region was reionized \citep[e.g.][]{daloisio15}. These effects cannot be fully captured in a homogeneous UVB model, but are discussed in detail based on radiative transfer simulations in \citet{keating17}.

In the middle panel of Fig.~\ref{fig:temp}, the temperature measurements based on the curvature of the Lyman-$\alpha$ forest were rescaled to mean cosmic density. A more direct comparison to the data of \citet{becker11} is possible by extracting mock Lyman-$\alpha$ forest spectra from cosmological hydrodynamical simulations, computing the curvature of the simulated spectra and converting it to a temperature at a characteristic density $\bar{\Delta}(z)$ with the same method as in \citet{becker11}. The results are presented in Fig.~\ref{fig:temp_Delta}. To give some idea of the accuracy of the curvature method for measuring temperatures, we also show the IGM temperature at density $\bar{\Delta}(z)$ as directly measured from the simulation with the fiducial model (shown is the mode of the temperature distribution of gas particles at that density). It is overall in good agreement with the temperature inferred from the curvature of the spectra with deviations $\Delta T(\bar{\Delta}(z)) \lesssim 2000 \mathrm{K}$. For our fiducial model, $T(\bar{\Delta}(z))$ computed from the spectra is in excellent agreement with the observational constraints for $z\lesssim 4$, while being slightly high above that. The latter suggests that there might be a mild tension between a late reionization finishing at $z\sim6$ and the IGM temperature measurements at $z=4.4$ and $4.8$. In particular models, with a harder AGN-like spectrum and correspondingly higher IGM temperatures are more strongly disfavoured. This will be discussed in more detail in Sec.~\ref{sec:agn_models}. 

\subsection{Lyman-$\alpha$ forest effective optical depths}
\label{sec:lya_taueff}

\begin{figure*}
\centerline{\includegraphics[width=\linewidth]{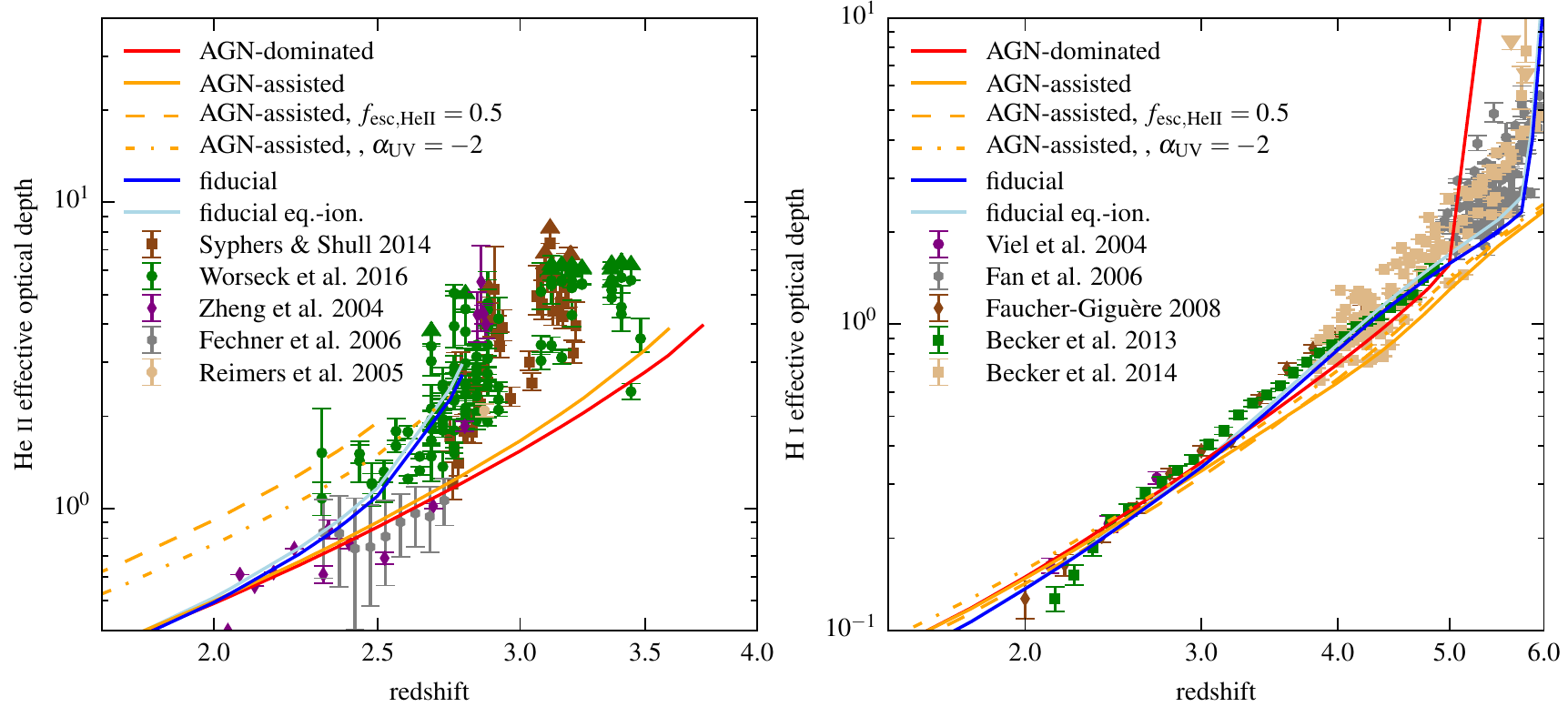}}
\caption{Effective optical depth of the \HeII (left panel) and \HI (right panel) Lyman-$\alpha$ forest for simulations based on our different UVB models as a function of redshift. Results are only plotted after percolation, since we do not expect a homogeneous UVB model to give sensible effective optical depths during a patchy reionization.}
\label{fig:tau_eff}
\end{figure*}

Another important property that simulations of the IGM should reproduce is the opacity of the Lyman-$\alpha$ forest, which is directly affected by the assumed UVB, but also sensitive to the IGM temperature. Fig.~\ref{fig:tau_eff} shows the effective optical depth defined by $\tau_{\rm eff} \equiv -\ln \langle F \rangle$, where $\langle F \rangle$ is the mean transmitted flux fraction. This is displayed both for the \HI and \HeII Lyman-$\alpha$ forest and for the different UVB models as a function of redshift. We show results only out to the redshift at which the reionization of the relevant species ends as homogeneous UVB models will underpredict residual transmission in ionized regions during inhomogeneous reionization. Our fiducial model is in good agreement with observations of both the \HI and \HeII Lyman-$\alpha$ forest. 

\subsection{Equivalent rates for codes assuming ionization equilibrium }
\label{sec:eq_rates}

Many state of the art cosmological hydrodynamical simulation projects, such as, e.g, Illustris \citep{vogelsberger2014} or Eagle \citep{schaye2015}, follow photoionization and photoheating under the simplifying assumption of ionization equilibrium in the presence of an external UV background. While this approach is numerically efficient, it is known that the ionization state of the IGM can deviate significantly from equilibrium, e.g., during reionization. Neglecting such deviations from equilibrium typically results in an earlier completion of reionization. It also artificially decouples the heat input from the change in ionization state. This biases the heating by reionization low \citep[see, e.g.,][]{puchwein15}. Some works compensate for this by introducing an ad hoc boost factor for the photoheating rates, e.g., during the epoch of \HeII reionization \citep[e.g.,][]{bolton2017}. In this work, we have avoided the problem altogether by using a more accurate full non-equilibrium treatment, i.e. rather than assuming ionization equilibrium, we compute the ionization state directly by integrating the ionization and recombination rate equations. Since many cosmological simulation codes still lack this functionality, we will in the following compute "equivalent-equilibrium" photoionization and photoheating rates that can be used in codes which assume ionization equilibrium. They will be chosen such that they allow equilibrium codes to recover the thermal and ionization history of the full non-equilibrium computation. We caution however that this strictly holds only at the mean cosmic density. The slope of the density-temperature relation will in general be somewhat different (see Fig.~\ref{fig:temp}).

Another aspect that is often neglected is the effect of the patchiness of reionization. Fully capturing this effect clearly requires radiative transfer simulations. With a homogeneous UVB model one can only attempt to recover the correct average ionization state and temperature, while fully capturing the spatial fluctuations is not possible. In Appendix~\ref{sec:temp_modelling_patchy} we discuss that even capturing the correct average values is not straightforward. For example, a model with a homogeneous UVB that has the same average ionization state and temperature as an inhomogeneously ionized IGM will still have different mean recombination and collisional cooling rates. This primarily results in differences in the thermal evolution of the IGM during hydrogen reionization. Fortunately the differences after the end of hydrogen reionization, where we compare our models to observational data, are rather small. Nevertheless we have elected to calibrate our "equivalent-equilibrium rates" on the $\Gamma = 10^{-14} \mathrm{s}^{-1}$ patchy reionization model presented in Appendix~\ref{sec:temp_modelling_patchy}. This results in a more accurate thermal evolution at $z \gtrsim 6$, while the temperatures at $z \lesssim 6$ are almost identical to those obtained from the non-equilibrium calculation with the homogeneous fiducial UVB (see Fig.~\ref{fig:Q}).

The "equivalent-equilibrium rates'' were computed as follows. We use the temperature and ionized fraction from the $\Gamma = 10^{-14} \mathrm{s}^{-1}$ patchy reionization model as an input. At each redshift, we first compute the \HI, \HeI and \HeII photoionization rates needed for gas that is in ionization equilibrium to have the same ionized fractions as the average of the patchy model. In this step, we assume the gas is at mean cosmic baryon density and we use the average temperature of the patchy model for computing the required recombination and collisional ionization rates. Using again the average ionized fractions and temperatures from the patchy model, we next compute all cooling rates at each redshift. We then compare the rate of temperature change expected from these cooling rates with the evolution of the average temperature in the patchy model. The difference between the two gives us the amount of photoheating that is needed for matching the evolution of the patchy model. The only remaining task is then to decide what fraction of the required amount of photoheating is contributed by each ion species. For this, we keep the ratio of the photoheating rates the same as in the fiducial (non-equilibrium) UVB model, i.e. we simply rescale the photoheating rates for \HI, \HeI and \HeIII all by the same redshift-dependent factor.

A full table of the ``equivalent-equilibrium'' photoheating and photoionization rates is given in Appendix~\ref{sec:rate_tables}. These rates are supposed to be used for following photoheating and photoionization in cosmological simulations with codes that assume ionization equilibrium, as, e.g., most versions of the \textsc{gadget} simulation code as well as many other cosmological simulation codes do.

\section{Models with a large AGN contribution}
\label{sec:agn_models}

The claim of a significant population of previously undetected faint AGN at $4 \lesssim z \lesssim 6.5$ by \citet{giallongo15} has sparked new interest in scenarios in which AGN contribute significantly to the ionizing photon budget during hydrogen reionization or at least at its tail end. MH15 have suggested that AGN could potentially be the main drivers of \HI reionization. \citet{chardin2015,chardin2017} have proposed them as a source of the Lyman-$\alpha$ forest opacity fluctuations on large ($50 h^{-1} \mathrm{cMpc}$) scales observed by \citet{becker15}.

UVB models using our new method of treating the effective opacity allow a more accurate prediction of the thermal evolution of the IGM after \HI and during \HeII reionization. In the following, we use this method to compute the UVB in scenarios with a large AGN contribution. With the derived photoionization and photoheating rates, we then simulate the thermal and ionization evolution of the IGM and compare it to observations. This allows us to put constraints on the ionizing emission from AGN. Except where specifically stated we use a full non-equilibrium ionization calculation. 

We will initially consider two scenarios with increased ionizing emission from AGN: i. As an upper extreme, our `AGN-dominated' model assumes the AGN ionizing emissivity from MH15, which at $z > 5$ was chosen to have the largest AGN emissivity still consistent with the \citet{giallongo15} measurements. In contrast to MH15, we, however, use our new treatment of the opacity for ionizing photons (see Sec.~\ref{sec:uvb_modelling}) when computing the UVB. The UVB in this model is completely dominated by AGN. ii. A model with an AGN ionizing emissivity identical to MH15 for $z < 4.75$, but with a much steeper fall-off for $z > 4.75$ (chosen simply as a straight line through the two higher redshift \citet{giallongo15} data points in Fig.~\ref{fig:eps912}). This model additionally includes stellar sources but with a lower emissivity than our fiducial model for $z \lesssim 7$ (adopting the SFRD of \citealt{madaufragos17}, $f_{\mathrm esc,0}=1.7\times10^{-7}$, $\alpha_{\mathrm esc}=7.2$, $f_{\mathrm esc,max}=0.5$).  Their ionizing emission nevertheless exceeds that of the AGN for $z \gtrsim 5$. Since AGN contribute to but do not dominate \HI reionization in this model, we call this the `AGN-assisted' model. This amount of ionizing emission from AGN may still be sufficient to explain the observed large-scale Lyman-$\alpha$ forest opacity fluctuations at $z \sim 5.4 - 5.6$ \citep[see, e.g.,][]{chardin2015,chardin2017}. The emissivity evolutions of both models are illustrated in Fig.~\ref{fig:eps912} and compared to our fiducial model and observational constraints. We will find below that both the AGN-dominated and AGN-assisted models tend to reionize \HeII too early. We will then discuss two further variations of the AGN-assisted model with a reduced emission above the \HeII ionization threshold, which aim to mitigate this. 

The top panel of Fig.~\ref{fig:temp} displays the redshift evolution of the \HII and \HeIII fractions predicted in our models with a larger AGN contribution to reionization. For each UVB model, we simulate the thermal and ionization state of the IGM. In the AGN-dominated model, \HI reionization ends at $z\sim5.6$, which is somewhat late compared to observational constraints. In contrast, \HeII reionization completes very early, at $z\sim3.9$. This is a consequence of the harder ionizing spectrum of AGN compared to stars. It also results in tension with the observed effective optical depth of the \HI and \HeII Lyman-$\alpha$ forest, as shown in Fig.~\ref{fig:tau_eff}. The AGN-dominated model overpredicts the \HI effective optical depth for $z \gtrsim 5$, while the \HeII effective optical depth is underpredicted for $z \gtrsim 2.8$. To complete \HI reionization earlier in an AGN-dominated scenario, we could further increase the AGN emissivity. This would, however, result in even stronger tension with the observed \HeII Lyman-$\alpha$ optical depth.

The small time-lag between \HI and \HeII reionization in the AGN-dominated model also results in a thermal evolution of the IGM that is quite different from our fiducial model in which stars drive \HI reionization. As shown in the middle panel of Fig.~\ref{fig:temp}, there are no separate peaks in the temperature at the end of \HI and \HeII reionization, instead there is a single broad temperature peak with $T_0 \approx 1.8 \times 10^4 \mathrm{K}$ at $z \sim 4-5$. The high temperatures at these redshifts are inconsistent with the IGM temperature measurements of \citet{becker11}. This is illustrated in more detail in Fig.~\ref{fig:temp_Delta},  which shows the IGM temperature at the over-density at which the curvature statistics used in \citet{becker11} is most sensitive. This avoids an extrapolation to the mean cosmic baryon density. In this more direct comparison, the IGM in the AGN-dominated model is also found to be much hotter than observed by \citet{becker11}. Combined with the disagreement with observed \HI and \HeII Lyman-$\alpha$ forest effective optical depths, this strongly disfavours the AGN-dominated model and more generally an \HI reionization that is driven predominantly by AGN with a hard spectrum.

Our AGN-assisted model completes \HI reionization earlier at $z \sim 6.8$. This gives the IGM more time to cool after reionization. Our hope was that this would bring the model into better agreement with IGM temperature measurements at $z < 5$. However, as shown in the middle panel of Fig.~\ref{fig:temp} and in more detail in Fig.~\ref{fig:temp_Delta}, the predicted temperature is still significantly higher than measured by \citet{becker11}. \HeII reionization also still completes early, at $z \sim 3.7$, which results in an underprediction of the \HeII Lyman-$\alpha$ effective optical depth for $z \gtrsim 2.8$ (see Fig.~\ref{fig:tau_eff}). It, hence, seems that even a scenario in which AGN with a hard spectrum dominant over stars only once \HI reionization has ended is disfavoured by post-reionization Lyman-$\alpha$ forest observations. The main issue is that \HeII reionization happens too early, which results not only in underpredicting the \HeII effective optical depth, but also in too much photoheating of the IGM at $4 \lesssim z \lesssim 6$. Overall, these finding are in good agreement with the results of \citet{daloisio17}.

We will now attempt to mitigate this by considering versions of the AGN-assisted model that have less emission above the \HeII ionization threshold. We consider two scenarios for this reduction: i. So far we have assumed an escape fraction of 100 per cent for ionizing photons emitted by AGN. For faint AGN this could overpredict the emission. In particular, due to the approximately four times larger recombination rate coefficient of \HeIII compared to \HII and the lower number of \hbox{He~$\scriptstyle\rm II$}-ionizing photons, the \HeII column density in AGN hosts could be larger than the \HI column density. The optical depth for \HeII ionizing photons could then exceed that for photons between the \HI and \HeII ionization edges. To model this we take a version of the AGN-assisted model in which all of the photons between the \HI and \HeII ionization edges escape, but only a fraction $f_{\rm esc,HeII} = 0.5$ of the \HeII ionizing photons. ii. As the second case, we consider AGN with an intrinsically softer spectrum. A power-law spectrum with a slope of $\alpha_\mathrm{UV} = -2$ rather than $-1.7$ is used. The emissivity at 912~\AA\ is kept constant. This, hence, reduces the number of \HeII ionizing photons. Results for these two additional models are also shown in Figs.~\ref{fig:temp}, \ref{fig:temp_Delta} and \ref{fig:tau_eff}.

Overall, both AGN-assisted models with a reduced number of \HeII ionizing photons behave rather similarly. Since \HeII ionizing photons are less abundant than \HI ionizing photons and largely absorbed by helium, the redshift at which \HI reionization completes does not change much. In contrast \HeII reionization is, as expected, delayed compared to the original AGN-assisted model. It finishes at $z\sim2.7$ and $2.9$ in the $f_{\rm esc,HeII} = 0.5$ and $\alpha_\mathrm{UV} = -2$ models respectively. \HeII reionization thus ends at about the same time as in our fiducial model. It is, however, significantly more extended, starting already at $z\sim7$. This early onset of \HeII reionization prevents the IGM from cooling more strongly after the completion of \HI reionization. The IGM temperature is hence larger than in our fiducial model and in tension with the \citet{becker11} measurements for $z \gtrsim 3$. The \HI Lyman-$\alpha$ forest opacity is slightly low compared to the data in the AGN-assisted models with less \HeII ionizing emission. The \HeII forest opacity is in agreement with the data at the end of \HeII reionization, but overpredicted later, at $z \lesssim 2.5$. At these redshifts, the ionizing emission from AGN in the original AGN-assisted model is comparable to our fiducial model. Thus the reduction of \HeII ionizing photons in the $f_{\rm esc,HeII} = 0.5$ and $\alpha_\mathrm{UV} = -2$ models results in a too large \HeII fraction and an overprediction of the opacity of the \HeII Lyman-$\alpha$ forest. These mismatches with observed opacities and temperatures imply that even models in which AGN contribute less than stars to \HI reionization and start to dominate the UVB only at $z \lesssim 5$ are difficult to bring into agreement with constraints from the \HI and \HeII Lyman-$\alpha$ forest. This would require adding further degrees of freedom beyond a constant reduction of the \HeII ionizing emission, such as a redshift-dependent escape fraction for \HeII ionizing photons. It also means that the significant contribution to the ionizing emissivity at $z>5$ from rare sources suggested by \citet{chardin2015,chardin2017} would have to be due to a class of sources different from normal AGN, i.e., due to sources that do not significantly contribute to the \HeII ionizing emissivity.

\section{Summary and conclusions}
\label{sec:conclusions}

We have here presented a new synthesis model of the cosmic UV background that is consistent with a late hydrogen reionization ending at $z \simeq 6$ as favoured by recent observational data, such as measurements of the Thomson scattering optical depth towards the CMB, the abundance of high redshift \lya emitters, and the opacity fluctuations in the high-redshift \lya forest. Our new fiducial model adopts the most recent determinations of the star formation rate in galaxies, and of the AGN luminosity evolution across cosmic time, together with new observational constraints on the \HI absorber column density distribution. The major improvement is, however, a new treatment of the IGM opacity for ionizing photons, which is able to faithfully capture the transition from a neutral to an almost fully ionized IGM during cosmic reionization.

With this new UVB model, we follow photoheating and photoionization in cosmological simulations to predict the redshift evolution of the IGM ionization and thermal state. Other than in some widely used simulations codes, we do not assume ionization equilibrium, but use a more accurate full non-equilibrium treatment. We also test how reliable such predictions are when based on homogeneous UVB models, which are widely employed in cosmological hydrodynamical simulations. Finally, we study models with a larger AGN contribution at high redshift as claimed by some recent measurements. Our main findings are:

\begin{itemize}[leftmargin=5mm]

\item Previous synthesis models of the UVB resulted in reionization and thermal histories in simulations that are inconsistent with the assumed ionizing emissivities. Reionization occurred typically much too early (e.g., at $z \sim 13$ for a HM12 UVB). This discrepancy is fully resolved by our new treatment of the IGM opacity for ionizing photons. 

\item In our fiducial UVB model, star forming galaxies leaking $\lta 18\%$ of their intrinsic Lyman continuum radiation into the IGM drive \HI reionization and complete it at $z \sim 6.2$. AGN drive \HeII reionization, which finishes at $z\sim 2.8$.

\item Some of the effects caused by a reionization process that is actually patchy are clearly missed when using a homogeneous UVB, such as large scale spatial fluctuations in the temperature. The mean IGM temperature is nevertheless in remarkably good agreement with a patchy reionization model after the end of hydrogen reionization and throughout \HeII reionization. Comparisons to post-reionization temperature constraints based on Lyman-$\alpha$ data should thus not be significantly affected.

\item Due to the limited time available for cooling between the end of \HI and the onset of \HeII reionization, higher IGM temperatures are predicted at $2 \lesssim z \lesssim 6$ if reionization finishes as late as suggested by CMB and \lya emitter data.  The difference is largest at $z \sim 6$ where the temperatures are higher by a factor 1.5 or more than predicted by simulations with the HM12 UVB. 

\item The IGM temperatures predicted for our fiducial UVB model are in excellent agreement with observational constraints at $z \lesssim 4$, while being slightly high compared to (somewhat uncertain) data at higher redshifts.

\item The mean transmissions of the \HI and \HeII \lya forests predicted with our fiducial UVB model are also in good agreement with observational constraints after the reionization of the relevant ion species. \HeII and \HI \lya opacities during the patchy reionization of these ion species are not expected to be realistically predicted with a homogeneous UVB model.

\item In models with a significant AGN contribution to \HI reionization, \HeII reionization typically occurs too early due to the harder ionizing spectrum of the AGN with more emission above the \HeII ionization edge. This results in an underprediction of the opacity of the \HeII Lyman-$\alpha$ forest and an overprediction of the IGM temperature at $z \gtrsim 2.8$. We have attempted to mitigate this by assuming that only a fraction of the \HeII ionizing photons escape from AGN, as well as by assuming a softer intrinsic AGN spectrum. While this results in a later end of \HeII reionization, comparable to our fiducial model, tensions with the measured IGM temperature and the \HeII Lyman-$\alpha$ forest opacity evolution persist.

\end{itemize}

Since not all simulation codes are able to follow non-equilibrium effects, in addition to the photoionization and photoheating rates of our new fiducial UVB model, we  also provide \textit{equivalent-equilibrium} rates that give a similar thermal and ionization evolution
in simulations with codes assuming ionization equilibrium.

The predictions of the thermal and ionization state of the IGM with numerical simulations have got to the state where comparison with \lya forest data poses severe constraints to any possible model of the cosmic UV background. While our fiducial model is in good agreement with the data our findings disfavour a significant contribution of normal AGN with a hard spectrum to \HI reionization.

\section*{Acknowledgements}

We thank James Bolton, Laura Keating and Girish Kulkarni for helpful discussions. EP acknowledges support by the Kavli Foundation. This work was further supported by the ERC Advanced Grant 320596 ``The Emergence of Structure during the epoch of Reionization". This work used the DiRAC (www.dirac.ac.uk) systems: Data Analytic at the University of Cambridge [funded by BIS National  E-infrastructure  capital  grant  (ST/K001590/1), STFC  capital  grants  ST/H008861/1  and  ST/H00887X/1, and   STFC   DiRAC   Operations   grant   ST/K00333X/1] and Data Centric at Durham University [funded by a BIS National E-infrastructure capital grant ST/K00042X/1, STFC capital grant ST/K00087X/1, DiRAC Operations grant ST/K003267/1 and Durham University]. DiRAC is part of the National E-Infrastructure. 


\bibliographystyle{mnras}
\bibliography{mybib}

\appendix

\section{Homogeneous vs patchy photoheating}
\label{sec:temp_modelling_patchy}

In cosmological hydrodynamical simulations photoheating is typically followed with a homogeneous UVB, e.g.,
with the widely used HM12 or \citet{faucher09} models. This approach clearly neglects some of the effects caused by a patchy
reionization process. Such simulations, hence, not only miss the patchy distribution of
neutral and ionized gas during the epoch of reionization, but also fail to capture IGM temperature fluctuations originating
from different regions being reionized at different redshifts. These fluctuations fade slowly and persist for some period after the
end of reionization \citep[e.g.,][]{daloisio15,keating17}. The hope is that at least (some suitable) average
temperature of the IGM is reasonably well captured in such simple simulations with a homogeneous UVB. In this appendix,
we will test some aspects of this assumption. In particular, we will study the effect on $T_0$, the temperature of the IGM at mean cosmic baryon density.  

A number of quantities that affect the average IGM temperature could potentially change when 
following photoheating with a homogeneous UVB compared to a more accurate treatment like, e.g.,
multi-frequency radiative transfer. This includes the: 
\begin{enumerate}[leftmargin=5mm]

\item \textit{redshift evolution of the ionized fraction} - Clearly a different time
evolution of the ionized fraction would affect both heating and cooling rates and alter
the thermal history of the IGM. With the new treatment of the mean free path presented
in Sec.~\ref{sec:uvb_modelling} such discrepancies in the time evolution of the ionized fraction are largely
removed. 
 
\item \textit{number of ionization events} - This number depends on how many recombinations
there are during the epoch of reionization, which result in some atoms being ionized
multiple times and hence conversion of more UV energy into thermal energy of the IGM.
How this differs between a homogeneous and a patchy reionization model will be discussed
in more detail below.

\item \textit{mean excess energy per ionization event} - The frequency dependence of
the photoionization cross section results in spectral filtering by residual
neutral gas between the ionizing source and a region of the IGM that is about to be ionized.
This affects the hardness of the spectrum that neutral hydrogen atoms at the edge of ionized bubbles
see and hence the excess energy that is available for heating there. Spectral filtering is included
in the calculations of the homogeneous UVB models, but may not capture all details of this process correctly.
Studying this in detail is, however, beyond the scope of this work as it would require full multi-frequency
radiative transfer simulations.

\item \textit{amount of cooling of the IGM} - During a patchy reionization the spatial distribution
of different ion species and of the IGM temperature is very inhomogeneous. Since many radiative cooling processes
stem from two-body interactions, the clustering of the ion species affects the volume-averaged cooling rates
even when their mean densities are the same as in a homogeneous model. We will investigate below how this
affects the dominant cooling processes.

\end{enumerate}

To investigate in particular points (ii) and (iv) in more detail, we compare the photoionization and photoheating by a
homogeneous UVB to a simple patchy reionization scenario. In the latter, we assume that the ionized gas is not smoothly
distributed but concentrated in ionized bubbles whose volume fraction is the same as the ionized fraction in the
homogeneous model. More precisely, we split the IGM into 1000 different regions that reionize almost instantaneously at different redshifts.
For each reionization redshift we compute the fraction of the volume that reionizes there. The volume fractions are computed
such that the evolution of the ionized fractions for both hydrogen and helium when volume averaged over all regions is the same as
in the homogeneous fiducial UVB model and thus also consistent with empirical emissivities. Calculating the temperature evolution by directly integrating all relevant rate equations (also allowing for non-equilibrium effects)
for each region then allows us to compare the volume-averaged
temperature evolution of patchy and homogeneous reionization models that have the same average ionized fractions as a function of
redshift. For simplicity we neglect correlations between reionization redshift and density and instead assume that all regions
are at mean cosmic density. We hence study essentially the effect of patchy reionization on $T_0$. The individual regions are reionized
by suddenly turning on an ionizing background with a chosen photoionization rate at the desired reionization redshift. The mean excess
energy per ionization is assumed to be the same as in the homogeneous fiducial UVB model.

\begin{figure}
\centerline{\includegraphics[width=\linewidth]{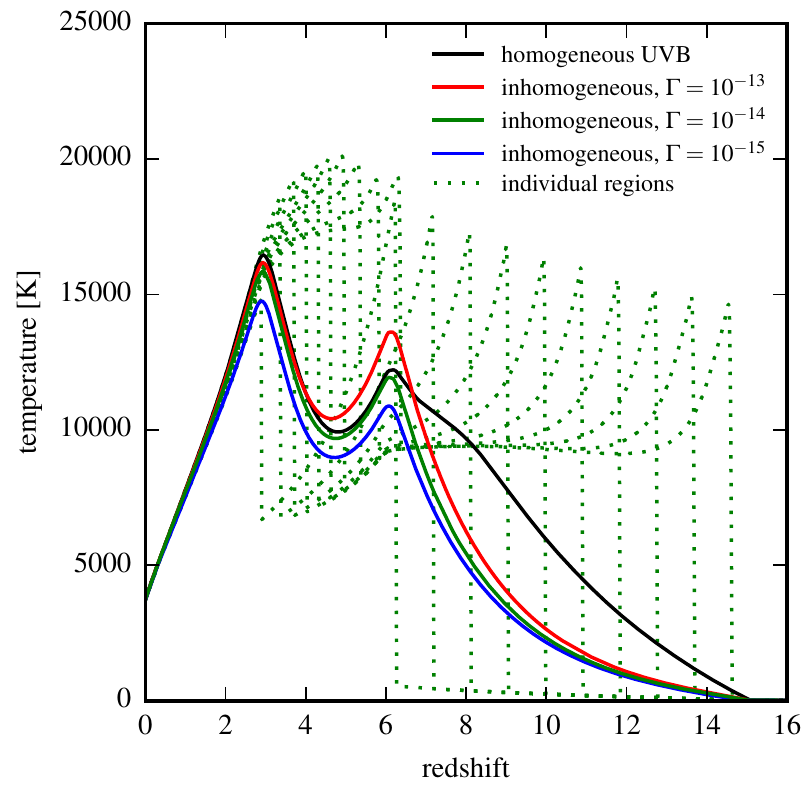}}
\caption{Effect of patchy reionization on the IGM temperature. The \textit{black, solid} curve shows
the temperature as a function of redshift when photoheating is followed with our fiducial homogeneous UVB model.
The \textit{green, dotted} curves show the thermal evolution of a subset of 10 of the regions with different reionization
redshifts. The \textit{green, solid} curve shows the volume average over all regions and can be compared to the homogeneous model.
The \textit{red, solid} curve shows the volume average for the patchy reionization model when the individual regions are
reionized more quickly, i.e., with a photoionization rate of $\Gamma=10^{-13} \mathrm{s}^{-1}$ rather than $\Gamma=10^{-14} \mathrm{s}^{-1}$. The \textit{blue, solid} curve shows results for a slower reionization with $\Gamma=10^{-15} \mathrm{s}^{-1}$. Overall the homogeneous and patchy models are in good agreement at $z \lesssim 6$ where most observational constraints on the IGM temperature fall.}
\label{fig:test_homog_temp}
\end{figure}

Fig.~\ref{fig:test_homog_temp} displays the thermal evolution of the IGM temperature in the homogeneous and patchy reionization models.
For the latter we consider three versions. One in which the individual regions are reionized rather quickly (with a photoionization rate of
$\Gamma=10^{-13} \mathrm{s}^{-1}$, corresponding to a time scale of $\sim3 \times 10^5 \mathrm{yr}$, the same value is used for all species), one intermediate model ($\Gamma=10^{-14} \mathrm{s}^{-1}$), and one in which the reionization of the individual regions happens more slowly ($\Gamma=10^{-15} \mathrm{s}^{-1}$). The \textit{dotted} curves show the thermal evolution for a subset of 10 of the individual regions in the $\Gamma=10^{-14} \mathrm{s}^{-1}$ model and illustrate the significant spatial scatter in the temperature during the epochs of \HI and \HeII reionization.
The volume average is, however, reasonably well reproduced by the homogeneous model after the end of hydrogen reionization where
most observational constraints fall. This suggests that using such simple homogeneous models in cosmological hydrodynamical simulations, while 
not being ideal, may be acceptable for many purposes, in particular when averaging the results over sufficiently large volumes and focusing primarily on $z \lesssim 6$. Note that our equivalent-equilibrium rates presented in Sec.~\ref{sec:eq_rates} take also the somewhat lower temperatures in the patchy model at $z \gtrsim 6$ into account, i.e., these rates were chosen such that the thermal evolution of the $\Gamma=10^{-14} \mathrm{s}^{-1}$ patchy model is followed. 

\begin{figure}
\centerline{\includegraphics[width=\linewidth]{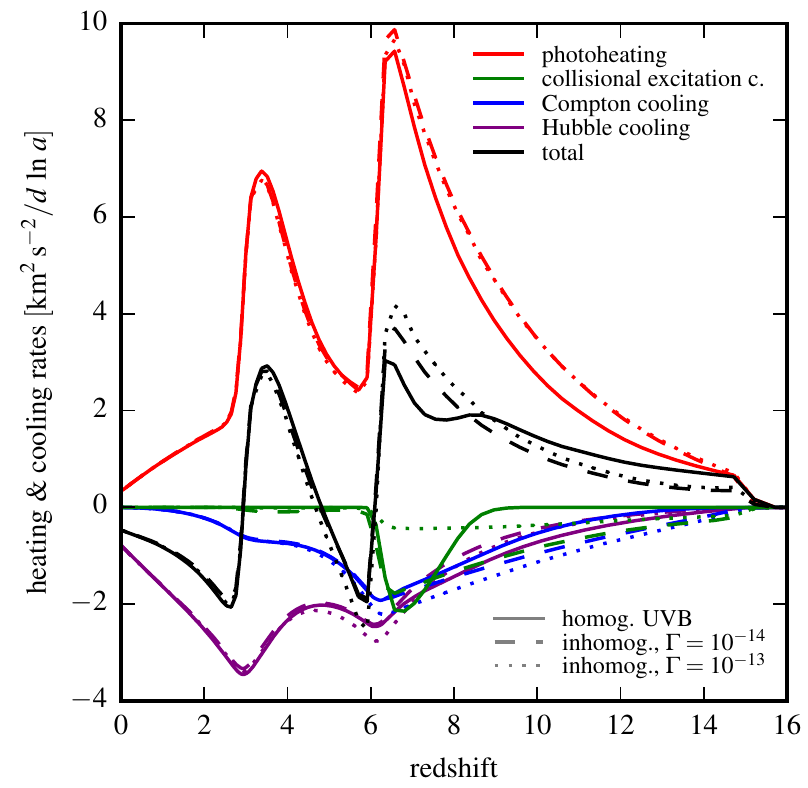}}
\caption{Effect of patchy reionization on the heating and cooling rates (per unit mass) of the IGM. 
Positive values correspond to heating, while negative values denote cooling.
\textit{Solid} lines show results for a calculation assuming photoheating and ionization by a homogeneous UVB. \textit{Dashed} and \textit{dotted} curves
display volume-averaged rates in the patchy reionization models with
$\Gamma=10^{-13} \mathrm{s}^{-1}$ and $\Gamma=10^{-14} \mathrm{s}^{-1}$, respectively. Subdominant processes like
collisional ionization, recombination and free-free cooling are not plotted for clarity. The curves for photoheating and collisional excitation
cooling have been slightly smoothed to remove residual noise from the finite number of regions used. The $\Gamma=10^{-15} \mathrm{s}^{-1}$ model is not shown for clarity.}
\label{fig:test_homog_heat}
\end{figure}

To understand differences between the patchy and homogeneous models during the epoch of hydrogen reionization better, it is illustrative
to look at the most important individual heating and cooling processes and how they are changed by a simplified reionization with a homogeneous UVB.
This is shown in Fig.~\ref{fig:test_homog_heat}. 

The volume-averaged photoheating in the patchy models is somewhat higher than in the homogeneous models
during hydrogen reionization. This can be understood by considering that both free electrons and \HII ions are concentrated within ionized bubbles
in the patchy model (as they should be), while they are artificially spread out over the whole volume in the homogeneous model. The latter suppresses the number of
recombinations and hence the number of atoms that are photoionized multiple times. This in turn reduces the associated photoheating. In the patchy models
the amount of photoheating is largely independent of how quickly the individual regions are reionized. 

Collisional excitation cooling is dominated by free electrons interacting with neutral hydrogen atoms. During patchy reionization this happens
primarily near the ionization fronts, i.e., in our model only in regions which are currently photoionized by a UVB that has been recently switched on.
In the homogeneous model instead, the artificial mixing of the different species allows collisional excitation cooling to happen everywhere. This can potentially
boost this process, which is indeed happening for $6 \lesssim z \lesssim 8$. Collisional excitation cooling is, however, also strongly temperature
dependent. It increases, e.g., by almost five orders of magnitude between 5,000K and 10,000K. Thus, at higher redshift, the homogeneous model is
simply too cold and the patchy models exhibit more collisional excitation cooling. The amount of such cooling is expected to depend on how quickly
ionization fronts propagate or in the context of our simple patchy model on how quickly the individual regions are reionized. As expected there is less collisional
excitation cooling when adopting a larger photoionization rate. This dependence means that there is some modelling uncertainty in this process.
We have chosen to present a model with $\Gamma=10^{-13} \mathrm{s}^{-1}$ as this is consistent with the mean photoionization rate found within ionized bubbles
in radiative transfer simulations at these redshifts \citep{chardin2017}. Since the time-averaged photoionization over the period in which the
ionization front passes is likely lower than that, we also present results for models with $\Gamma=10^{-14} \mathrm{s}^{-1}$ and $10^{-15} \mathrm{s}^{-1}$.
Despite a factor of 100 difference in the assumed photoionization rate, the thermal evolution in the three models differs by $\lesssim 25 \%$, suggesting that this modelling uncertainty is reasonably well under control.

Compton cooling is not directly affected by the clustering of the free electrons. As the CMB photons are uniformly distributed, it depends only on the
number of free electrons, which by construction is the same as in the homogeneous model, and their temperature. Hence, the differences between the models
are solely caused by the temperature dependence of Compton cooling. In the patchy models the electrons reside in ionized bubbles which are typically hotter than the
IGM in the homogeneous model. This results in suppressed Compton cooling in the latter model.

Finally, the Hubble cooling rate depends only on the temperature and hence simply reflects the differences in the thermal evolution that is caused by the 
other effects discussed above. The total heating rate explains the effects seen in Fig.~\ref{fig:test_homog_temp}. In particular the ``knee'' in the
thermal evolution of the homogeneous model at $z\approx8$ is caused by the rather sudden onset of efficient collisional excitation cooling in that model. This ``knee'' should thus be considered an artefact of the homogeneous treatment.

In principle, many of the mechanisms discussed above apply in similar form during \HeII reionization. Their effects on the thermal evolution are
however strongly limited there. This can be understood by considering that the free electron density and temperature differ only by factors $\sim 1.07$ and $\sim2$ between
\HeII and \HeIII regions, compared to several orders of magnitude between \HI and \HII regions.
Hence, the homogeneous model is able to reproduce the average thermal evolution of the patchy models well after the end of hydrogen reionization.

\section{Estimating the mean excess energy in the complete, local absorption and transparent IGM limits}
\label{sec:energy_limits}

In Fig.~\ref{fig:rates}, we have presented the mean excess energy (beyond the ionization threshold) of \HI and \HeII ionizing photons in our fiducial UVB model. This excess energy is available for photoheating the IGM. We have compared it to the complete local absorption (optically thick) and transparent IGM (optically thin) limits, as well as to the HM12 model.  Here, we provide the details of how these excess energies were computed.

The energies shown for our fiducial and the HM12 model are simply given by the ratio, $\mathcal{H}_i /  \Gamma_i$, of the photoheating rate,
\begin{equation}
 \mathcal{H}_i = \int_{\nu_i}^\infty \textrm{d} \nu \, \frac{4 \pi J_\nu}{h \nu} \, h (\nu - \nu_i) \, \sigma_i(\nu),
\label{eq:Hi}
\end{equation}
to the photoionization rate,
\begin{equation}
  \Gamma_i = \int_{\nu_i}^\infty\textrm{d} \nu \, \frac{4 \pi J_\nu}{h \nu} \, \sigma_i(\nu),
\label{eq:Gammai}
\end{equation}
where $h$ is Planck's constant, the subscript $i$ denotes the relevant ion species, $h \nu_i$ its ionization energy,
and $\sigma_i(\nu)$ its photoionization cross section. $J_\nu$ is the background intensity as defined in Eq.~(\ref{eq:Jnu}).

The excess energy in the transparent IGM limit, which is approached at late times, was computed by neglecting the intergalactic opacity when computing $J_\nu$, i.e. by setting $\bar{\tau} = 0$ in Eq.~(\ref{eq:Jnu}). The excess energy is then computed from this optical thin $J_\nu$ by  using Eqs.~(\ref{eq:Hi}) and (\ref{eq:Gammai}) and again deriving the ratio $\mathcal{H}_i / \Gamma_i$.

In the complete local absorption limit, we calculate the mean excess energy by
\begin{equation}
  E_{\rm excess, thick} = \frac{\int_{\nu_i}^{\nu_{\rm max}}\ \textrm{d} \nu \, \frac{4 \pi \epsilon_\nu}{h \nu} \, h (\nu - \nu_i) \, f_i(\nu)} 
      {\int_{\nu_i}^{\nu_{\rm max}} \textrm{d} \nu \, \frac{4 \pi \epsilon_\nu}{h \nu} \, f_i(\nu)}.
\end{equation}
Since we assume local absorption, we directly use the emitted spectrum $\epsilon_\nu$ for computing the excess energy. The factor
\begin{equation}
  f_i(\nu) = \frac{n_i \sigma_i(\nu)} {\sum n_j \sigma_j(\nu)},
\end{equation}
where the sum extends over the species \hbox{H~$\scriptstyle\rm I$}, \HeI and \hbox{He~$\scriptstyle\rm II$}, is used to estimate the fraction of the absorption by each ion species. Here, $n_i$ is the mean number density of ion $i$ at the considered redshift, which is obtained from the simulation prediction of the ionized fraction in the fiducial UVB model as shown in the top panel of Fig.~\ref{fig:Q}. For \hbox{He~$\scriptstyle\rm II$}, the mean excess energy in the complete local absorption limit depends somewhat on the assumed upper limit of the photon frequency, $\nu_{\rm max}$. In particular, it depends on whether or not the $\sim30$~keV bump in the AGN spectrum (see HM12) is included. As even a completely neutral universe is no longer optical thick to such high-energy photons, we choose $\nu_{\rm max} = 1 \, \mathrm{keV} / h$ as the upper limit of the integration.

\section{Full cosmological simulations vs one-cell calculations}
\label{sec:cosmo_vc_one-cell}

\begin{figure}
\centerline{\includegraphics[width=\linewidth]{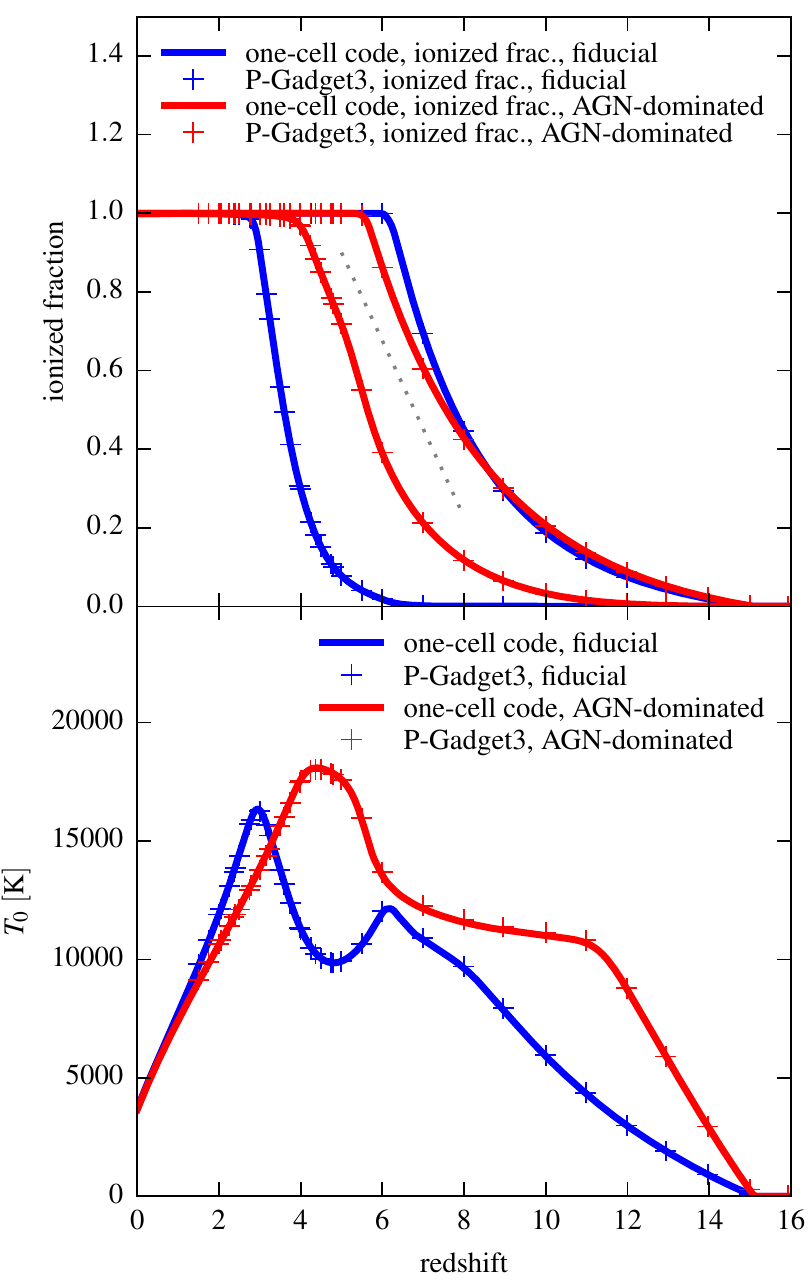}}
\caption{Comparison of the ionization (top panel) and thermal (bottom panel) evolution of the IGM predicted by \textsc{p-gadget3} with the non-equilibrium ionization solver to results from our lightweight one-cell code. Results are shown for our fiducial and AGN-dominated UVB models. In the top panel, the curves to the left of the gray dotted line show \HeIII fractions, while those to the right show \HII fractions. For both models, the predictions with the different codes are in excellent agreement.}
\label{fig:one-cell_vs_sim}
\end{figure}

For quickly exploring the parameter space of UVB models, e.g., regarding the choice of galactic escape fraction, or suitable AGN emission models, we have used a lightweight one-cell code. It follows the thermal and ionization evolution of a single gas cell at mean cosmic baryon density that is exposed to the considered UVB. This only takes a few seconds per UVB model. The ionization, recombination, heating and cooling rates assumed in this code are identical to those used in our non-equilibrium ionization solver in the full cosmological simulations with \textsc{p-gadget3} (see \citealt{puchwein15} for details). Non-equilibrium effects are taken into account in the one-cell code as well.   

In Fig.~\ref{fig:Q} and in the upper two panels of Fig.~\ref{fig:temp}, we show predictions of the thermal and ionization evolution of the IGM. These are based on this one-cell code rather than on full cosmological simulations. This has the advantage that we get better time resolution then for the cosmological simulations. For the latter, saving a comparably large number of full snapshot files would be expensive in terms of disk space. Fig.~\ref{fig:one-cell_vs_sim} demonstrates that the thermal and ionization evolutions computed with the one-cell code are in excellent agreement with the full cosmological simulations. In the bottom panel, the temperature at mean density is shown for our fiducial and AGN-dominated models. For the cosmological simulations this was computed as the median temperature of all gas particles with densities between $0.95$ and $1.05$ times the mean cosmic baryon density. The one-cell code directly predicts the temperate at mean density. The two predictions are in excellent agreement. The top panel shows the \HII and \HeIII fractions. For the simulations this was computed as a mass-weighted average of the ionization fraction of all (no matter what their density) gas particles. For the one-cell code, simply the ionization fraction of the single cell at mean density is shown. Again, the predictions with the different codes are in excellent agreement.

Note that the results shown in the bottom panel of Fig.~\ref{fig:temp}, as well as in Figs.~\ref{fig:temp_Delta} and \ref{fig:tau_eff} are based on full cosmological simulations as they require predictions of IGM properties beyond mean density. These simulations use a box with a side length of $10 \, h^{-1}$ comoving Mpc which is populated with $512^3$ gas and the same number of dark matter particles. The simulation were performed with a modified version of \textsc{p-gadget3} \citep[see][]{puchwein15}.

\section{Tables of the photoionization and photoheating rates}
\label{sec:rate_tables}

\subsection{Rates for our fiducial UVB model}

Table~\ref{tab:rates} provides photoionization and photoheating rates for \hbox{H~$\scriptstyle\rm I$}, \HeI and \HeII as a function of redshift for our fiducial UVB model with the new intergalactic opacity treatment. To get these, the UVB intensity, $J_\nu$ was computed as described in Sec.~\ref{sec:methods}. The rates were than obtained from $J_\nu$ using Eqs.~(\ref{eq:Gammai}) and (\ref{eq:Hi}). The redshift evolution of the rates for \HI and \HeII is displayed in Fig.~\ref{fig:rates}. Electronic versions of this table are available both on the journal and arXiv article web pages.

Note that in cosmological simulations, these rates should only be used if the simulation code performs a full non-equilibrium ionization calculation, i.e., if it directly integrates the ionization and recombination rate equations rather than assuming ionization equilibrium. In the latter case, we suggest to use the rates provided in the next sub-section.

\subsection{Equivalent-equilibrium rates for use in cosmological simulation codes which assume ionization equilibrium}

Table~\ref{tab:eq_rates} provides equivalent-equilibrium photoionization and photoheating rates for \hbox{H~$\scriptstyle\rm I$}, \HeI and \HeII as a function of redshift. Electronic versions of this table are available both on the journal and arXiv article web pages. For users of the \textsc{gadget} code a table of the rates in the TREECOOL file format used by \textsc{gadget} is also available there. Compared to the rates of the fiducial UVB model, we have adopted a somewhat finer redshift sampling for the equivalent-equilibrium rates to get smooth thermal evolutions even when assuming ionization equilibrium.

As discussed in Sec.~\ref{sec:eq_rates}, these rates were chosen such that simulation codes that assume ionization equilibrium are able to recover a similar evolution of the thermal and ionization state as full non-equilibrium ionization codes do with our fiducial UVB model. There are however some notable differences. First, while there will be good agreement at mean cosmic baryon density, the slope of the temperature-density relation will be somewhat different (see Fig.~\ref{fig:temp}). This is unavoidable since under the assumption of ionization equilibrium, photoionizations balance recombinations. Photoheating is thus too efficient at high densities, where more recombinations occur, during reionization. Second, while excellent agreement is found with non-equilibrium ionization predictions based on our fiducial UVB model for $z \lesssim 6$, there is some difference at higher redshifts. In particular, we have elected to compute the equivalent-equilibrium rates such that the thermal and ionization history of the $\Gamma=10^{-14} \mathrm{s}^{-1}$ patchy reionization model presented in Appendix~\ref{sec:temp_modelling_patchy} is followed. This allows us to capture some effects that would normally be missed when treating photoheating with a homogeneous UVB, such as getting the correct volume-averaged collisional excitation cooling rates. This results in slightly lower temperatures at $z \gtrsim 6$ (see Fig.~\ref{fig:Q}).

Overall, our equivalent-equilibrium model when used in simulation codes assuming ionization equilibrium will do slightly better on the IGM temperature
at mean density at $z \gtrsim 6$ compared to our fiducial UVB model evolved with a non-equilibrium solver. The temperature at mean density for $z \lesssim 6$ will be almost identical. The fiducial UVB used with a non-equilibrium ionization and heating solver will, however, predict a more accurate slope of the temperature-density relation. We expect the latter to be more important for many applications and hence suggest to use the full fiducial UVB model in codes that support a non-equilibrium treatment.

\begin{table*}
\begin{tabular}{lccccccc}
\hline
$z$ & $\Gamma_{\mathrm{HI}}$ & $\mathcal{H}_{\mathrm{HI}}$ & $\Gamma_{\mathrm{HeI}}$ & $\mathcal{H}_{\mathrm{HeI}}$ & $\Gamma_{\mathrm{HeII}}$ & $\mathcal{H}_{\mathrm{HeII}}$ \\
& [s$^{-1}$] & [eV s$^{-1}$] & [s$^{-1}$] & [eV s$^{-1}$] & [s$^{-1}$] & [eV s$^{-1}$] \\
\hline
0.0 & 6.06e-14 & 2.28e-13 & 3.03e-14 & 2.49e-13 & 1.10e-15 & 2.08e-14 \\
0.0491 & 7.43e-14 & 2.81e-13 & 3.76e-14 & 3.06e-13 & 1.32e-15 & 2.50e-14 \\
0.101 & 9.08e-14 & 3.44e-13 & 4.63e-14 & 3.75e-13 & 1.59e-15 & 3.00e-14 \\
0.155 & 1.10e-13 & 4.19e-13 & 5.68e-14 & 4.58e-13 & 1.90e-15 & 3.59e-14 \\
0.211 & 1.34e-13 & 5.09e-13 & 6.94e-14 & 5.58e-13 & 2.26e-15 & 4.28e-14 \\
0.271 & 1.61e-13 & 6.15e-13 & 8.44e-14 & 6.76e-13 & 2.69e-15 & 5.08e-14 \\
0.333 & 1.94e-13 & 7.40e-13 & 1.02e-13 & 8.15e-13 & 3.19e-15 & 6.00e-14 \\
0.399 & 2.31e-13 & 8.85e-13 & 1.23e-13 & 9.77e-13 & 3.77e-15 & 7.07e-14 \\
0.468 & 2.75e-13 & 1.05e-12 & 1.46e-13 & 1.17e-12 & 4.43e-15 & 8.27e-14 \\
0.54 & 3.24e-13 & 1.24e-12 & 1.74e-13 & 1.38e-12 & 5.17e-15 & 9.63e-14 \\
0.615 & 3.81e-13 & 1.46e-12 & 2.05e-13 & 1.62e-12 & 6.00e-15 & 1.11e-13 \\
0.695 & 4.43e-13 & 1.70e-12 & 2.39e-13 & 1.89e-12 & 6.90e-15 & 1.28e-13 \\
0.778 & 5.13e-13 & 1.97e-12 & 2.77e-13 & 2.19e-12 & 7.88e-15 & 1.45e-13 \\
0.865 & 5.88e-13 & 2.26e-12 & 3.19e-13 & 2.51e-12 & 8.91e-15 & 1.64e-13 \\
0.957 & 6.69e-13 & 2.56e-12 & 3.63e-13 & 2.85e-12 & 9.97e-15 & 1.83e-13 \\
1.05 & 7.53e-13 & 2.88e-12 & 4.08e-13 & 3.20e-12 & 1.10e-14 & 2.01e-13 \\
1.15 & 8.38e-13 & 3.20e-12 & 4.54e-13 & 3.55e-12 & 1.19e-14 & 2.18e-13 \\
1.26 & 9.22e-13 & 3.52e-12 & 4.99e-13 & 3.88e-12 & 1.27e-14 & 2.33e-13 \\
1.37 & 1.00e-12 & 3.81e-12 & 5.39e-13 & 4.17e-12 & 1.32e-14 & 2.43e-13 \\
1.49 & 1.07e-12 & 4.06e-12 & 5.74e-13 & 4.41e-12 & 1.33e-14 & 2.46e-13 \\
1.61 & 1.13e-12 & 4.26e-12 & 6.01e-13 & 4.58e-12 & 1.30e-14 & 2.43e-13 \\
1.74 & 1.17e-12 & 4.42e-12 & 6.19e-13 & 4.68e-12 & 1.24e-14 & 2.33e-13 \\
1.87 & 1.20e-12 & 4.52e-12 & 6.29e-13 & 4.70e-12 & 1.15e-14 & 2.18e-13 \\
2.01 & 1.21e-12 & 4.55e-12 & 6.28e-13 & 4.63e-12 & 1.04e-14 & 1.98e-13 \\
2.16 & 1.21e-12 & 4.52e-12 & 6.17e-13 & 4.46e-12 & 8.92e-15 & 1.72e-13 \\
2.32 & 1.20e-12 & 4.43e-12 & 5.96e-13 & 4.20e-12 & 7.23e-15 & 1.42e-13 \\
2.48 & 1.17e-12 & 4.29e-12 & 5.66e-13 & 3.86e-12 & 5.30e-15 & 1.07e-13 \\
2.65 & 1.12e-12 & 4.10e-12 & 5.30e-13 & 3.45e-12 & 3.17e-15 & 6.99e-14 \\
2.83 & 1.07e-12 & 3.89e-12 & 4.93e-13 & 3.03e-12 & 1.21e-15 & 3.45e-14 \\
3.02 & 1.02e-12 & 3.69e-12 & 4.61e-13 & 2.75e-12 & 4.26e-16 & 1.58e-14 \\
3.21 & 9.68e-13 & 3.50e-12 & 4.33e-13 & 2.54e-12 & 2.10e-16 & 9.00e-15 \\
3.42 & 9.18e-13 & 3.32e-12 & 4.09e-13 & 2.38e-12 & 1.35e-16 & 6.17e-15 \\
3.64 & 8.73e-13 & 3.17e-12 & 3.90e-13 & 2.24e-12 & 9.56e-17 & 4.51e-15 \\
3.87 & 8.34e-13 & 3.04e-12 & 3.75e-13 & 2.15e-12 & 7.04e-17 & 3.38e-15 \\
4.11 & 8.02e-13 & 2.94e-12 & 3.65e-13 & 2.09e-12 & 5.30e-17 & 2.56e-15 \\
4.36 & 7.77e-13 & 2.86e-12 & 3.60e-13 & 2.05e-12 & 4.04e-17 & 1.94e-15 \\
4.62 & 7.58e-13 & 2.81e-12 & 3.59e-13 & 2.04e-12 & 2.98e-17 & 1.44e-15 \\
4.89 & 7.45e-13 & 2.79e-12 & 3.61e-13 & 2.04e-12 & 2.21e-17 & 1.06e-15 \\
5.18 & 7.32e-13 & 2.76e-12 & 3.63e-13 & 2.05e-12 & 1.66e-17 & 7.72e-16 \\
5.49 & 7.09e-13 & 2.68e-12 & 3.55e-13 & 2.00e-12 & 1.26e-17 & 5.58e-16 \\
5.81 & 6.09e-13 & 2.29e-12 & 2.97e-13 & 1.66e-12 & 9.88e-18 & 4.02e-16 \\
6.14 & 3.11e-15 & 1.86e-14 & 3.77e-15 & 3.21e-14 & 7.83e-18 & 2.82e-16 \\
6.49 & 7.29e-16 & 4.41e-15 & 8.94e-16 & 7.65e-15 & 3.45e-18 & 1.45e-16 \\
6.86 & 3.37e-16 & 2.05e-15 & 4.16e-16 & 3.55e-15 & 1.72e-18 & 8.03e-17 \\
7.25 & 2.15e-16 & 1.32e-15 & 2.68e-16 & 2.26e-15 & 9.34e-19 & 4.54e-17 \\
7.65 & 1.56e-16 & 9.58e-16 & 1.95e-16 & 1.63e-15 & 5.21e-19 & 2.56e-17 \\
8.07 & 1.20e-16 & 7.44e-16 & 1.52e-16 & 1.26e-15 & 2.94e-19 & 1.43e-17 \\
8.52 & 9.67e-17 & 6.02e-16 & 1.24e-16 & 1.02e-15 & 1.68e-19 & 7.95e-18 \\
8.99 & 7.96e-17 & 4.99e-16 & 1.03e-16 & 8.49e-16 & 9.97e-20 & 4.46e-18 \\
9.48 & 6.67e-17 & 4.21e-16 & 8.70e-17 & 7.19e-16 & 6.27e-20 & 2.58e-18 \\
9.99 & 5.66e-17 & 3.58e-16 & 7.43e-17 & 6.15e-16 & 4.28e-20 & 1.56e-18 \\
10.5 & 4.83e-17 & 3.08e-16 & 6.40e-17 & 5.31e-16 & 3.21e-20 & 1.01e-18 \\
11.1 & 4.15e-17 & 2.65e-16 & 5.52e-17 & 4.59e-16 & 2.59e-20 & 6.96e-19 \\
11.7 & 3.56e-17 & 2.27e-16 & 4.74e-17 & 3.94e-16 & 2.11e-20 & 5.00e-19 \\
12.3 & 3.06e-17 & 1.96e-16 & 4.08e-17 & 3.39e-16 & 1.78e-20 & 3.72e-19 \\
13.0 & 2.65e-17 & 1.69e-16 & 3.53e-17 & 2.93e-16 & 1.54e-20 & 2.82e-19 \\
13.7 & 2.30e-17 & 1.47e-16 & 3.06e-17 & 2.55e-16 & 1.33e-20 & 2.13e-19 \\
14.4 & 2.00e-17 & 1.28e-16 & 2.66e-17 & 2.21e-16 & 1.16e-20 & 1.58e-19 \\
15.1 & 1.74e-17 & 1.11e-16 & 2.32e-17 & 1.92e-16 & 1.01e-20 & 1.10e-19 \\
\hline
\end{tabular}
\caption{Photoionization and photoheating rates for \hbox{H~$\scriptstyle\rm I$}, \HeI and \HeII in our \textit{fiducial} cosmic UV background model as a function of redshift. When following photoheating and photoionization in cosmological simulations, these rates should only be used if the simulation code uses a full non-equilibrium ionization and heating solver. Electronic versions of this table are available both on the journal and arXiv article web pages. For simulation codes assuming ionization equilibrium use the rates provided in Table~\ref{tab:eq_rates}.}
\label{tab:rates}
\end{table*}

\begin{table*}
\begin{tabular}{lccccccc}
\hline
$z$ & $\Gamma_{\mathrm{HI}}$ & $\mathcal{H}_{\mathrm{HI}}$ & $\Gamma_{\mathrm{HeI}}$ & $\mathcal{H}_{\mathrm{HeI}}$ & $\Gamma_{\mathrm{HeII}}$ & $\mathcal{H}_{\mathrm{HeII}}$ \\
& [s$^{-1}$] & [eV s$^{-1}$] & [s$^{-1}$] & [eV s$^{-1}$] & [s$^{-1}$] & [eV s$^{-1}$] \\
\hline
0.0000 & 6.06e-14 & 2.28e-13 & 3.03e-14 & 2.49e-13 & 1.10e-15 & 2.08e-14 \\
0.0243 & 6.72e-14 & 2.54e-13 & 3.38e-14 & 2.77e-13 & 1.21e-15 & 2.29e-14 \\
0.0491 & 7.44e-14 & 2.82e-13 & 3.77e-14 & 3.08e-13 & 1.33e-15 & 2.51e-14 \\
0.0747 & 8.23e-14 & 3.13e-13 & 4.18e-14 & 3.40e-13 & 1.45e-15 & 2.75e-14 \\
0.1010 & 9.10e-14 & 3.46e-13 & 4.64e-14 & 3.77e-13 & 1.60e-15 & 3.01e-14 \\
0.1277 & 1.00e-13 & 3.82e-13 & 5.14e-14 & 4.16e-13 & 1.74e-15 & 3.30e-14 \\
0.1550 & 1.10e-13 & 4.21e-13 & 5.69e-14 & 4.60e-13 & 1.91e-15 & 3.61e-14 \\
0.1827 & 1.22e-13 & 4.64e-13 & 6.29e-14 & 5.08e-13 & 2.08e-15 & 3.94e-14 \\
0.2110 & 1.34e-13 & 5.11e-13 & 6.96e-14 & 5.61e-13 & 2.27e-15 & 4.30e-14 \\
0.2406 & 1.47e-13 & 5.62e-13 & 7.67e-14 & 6.17e-13 & 2.47e-15 & 4.69e-14 \\
0.2710 & 1.61e-13 & 6.18e-13 & 8.46e-14 & 6.79e-13 & 2.70e-15 & 5.10e-14 \\
0.3016 & 1.77e-13 & 6.78e-13 & 9.30e-14 & 7.46e-13 & 2.94e-15 & 5.55e-14 \\
0.3330 & 1.95e-13 & 7.44e-13 & 1.02e-13 & 8.19e-13 & 3.20e-15 & 6.03e-14 \\
0.3656 & 2.12e-13 & 8.13e-13 & 1.12e-13 & 8.97e-13 & 3.48e-15 & 6.55e-14 \\
0.3990 & 2.32e-13 & 8.89e-13 & 1.23e-13 & 9.82e-13 & 3.78e-15 & 7.10e-14 \\
0.4331 & 2.53e-13 & 9.69e-13 & 1.34e-13 & 1.07e-12 & 4.10e-15 & 7.69e-14 \\
0.4680 & 2.76e-13 & 1.06e-12 & 1.46e-13 & 1.18e-12 & 4.45e-15 & 8.31e-14 \\
0.5036 & 3.00e-13 & 1.15e-12 & 1.60e-13 & 1.28e-12 & 4.81e-15 & 8.97e-14 \\
0.5400 & 3.25e-13 & 1.25e-12 & 1.75e-13 & 1.39e-12 & 5.19e-15 & 9.68e-14 \\
0.5771 & 3.53e-13 & 1.35e-12 & 1.90e-13 & 1.50e-12 & 5.59e-15 & 1.04e-13 \\
0.6150 & 3.83e-13 & 1.47e-12 & 2.06e-13 & 1.63e-12 & 6.03e-15 & 1.12e-13 \\
0.6545 & 4.13e-13 & 1.58e-12 & 2.22e-13 & 1.76e-12 & 6.46e-15 & 1.20e-13 \\
0.6950 & 4.45e-13 & 1.71e-12 & 2.40e-13 & 1.90e-12 & 6.93e-15 & 1.29e-13 \\
0.7360 & 4.79e-13 & 1.84e-12 & 2.58e-13 & 2.05e-12 & 7.41e-15 & 1.37e-13 \\
0.7780 & 5.16e-13 & 1.98e-12 & 2.78e-13 & 2.20e-12 & 7.92e-15 & 1.46e-13 \\
0.8210 & 5.52e-13 & 2.12e-12 & 2.99e-13 & 2.36e-12 & 8.42e-15 & 1.55e-13 \\
0.8650 & 5.91e-13 & 2.28e-12 & 3.21e-13 & 2.53e-12 & 8.96e-15 & 1.65e-13 \\
0.9104 & 6.31e-13 & 2.42e-12 & 3.42e-13 & 2.69e-12 & 9.48e-15 & 1.75e-13 \\
0.9570 & 6.73e-13 & 2.58e-12 & 3.65e-13 & 2.87e-12 & 1.00e-14 & 1.84e-13 \\
1.0030 & 7.15e-13 & 2.74e-12 & 3.87e-13 & 3.04e-12 & 1.05e-14 & 1.93e-13 \\
1.0500 & 7.58e-13 & 2.90e-12 & 4.11e-13 & 3.23e-12 & 1.11e-14 & 2.03e-13 \\
1.0994 & 8.00e-13 & 3.06e-12 & 4.33e-13 & 3.40e-12 & 1.15e-14 & 2.11e-13 \\
1.1500 & 8.45e-13 & 3.23e-12 & 4.57e-13 & 3.58e-12 & 1.20e-14 & 2.20e-13 \\
1.2043 & 8.86e-13 & 3.39e-12 & 4.80e-13 & 3.75e-12 & 1.24e-14 & 2.27e-13 \\
1.2600 & 9.30e-13 & 3.55e-12 & 5.03e-13 & 3.92e-12 & 1.28e-14 & 2.35e-13 \\
1.3143 & 9.69e-13 & 3.70e-12 & 5.23e-13 & 4.06e-12 & 1.31e-14 & 2.40e-13 \\
1.3700 & 1.01e-12 & 3.85e-12 & 5.44e-13 & 4.21e-12 & 1.33e-14 & 2.45e-13 \\
1.4293 & 1.05e-12 & 3.98e-12 & 5.61e-13 & 4.33e-12 & 1.34e-14 & 2.47e-13 \\
1.4900 & 1.08e-12 & 4.13e-12 & 5.80e-13 & 4.49e-12 & 1.34e-14 & 2.50e-13 \\
1.5493 & 1.11e-12 & 4.21e-12 & 5.93e-13 & 4.55e-12 & 1.33e-14 & 2.47e-13 \\
1.6100 & 1.14e-12 & 4.31e-12 & 6.07e-13 & 4.63e-12 & 1.31e-14 & 2.46e-13 \\
1.6742 & 1.16e-12 & 4.39e-12 & 6.17e-13 & 4.69e-12 & 1.28e-14 & 2.41e-13 \\
1.7400 & 1.19e-12 & 4.48e-12 & 6.26e-13 & 4.74e-12 & 1.25e-14 & 2.36e-13 \\
1.8042 & 1.20e-12 & 4.53e-12 & 6.32e-13 & 4.75e-12 & 1.21e-14 & 2.28e-13 \\
1.8700 & 1.22e-12 & 4.58e-12 & 6.37e-13 & 4.76e-12 & 1.16e-14 & 2.21e-13 \\
1.9392 & 1.22e-12 & 4.60e-12 & 6.37e-13 & 4.73e-12 & 1.11e-14 & 2.11e-13 \\
2.0100 & 1.23e-12 & 4.64e-12 & 6.37e-13 & 4.72e-12 & 1.05e-14 & 2.02e-13 \\
2.0841 & 1.23e-12 & 4.58e-12 & 6.32e-13 & 4.59e-12 & 9.75e-15 & 1.87e-13 \\
2.1600 & 1.23e-12 & 4.58e-12 & 6.27e-13 & 4.52e-12 & 9.02e-15 & 1.74e-13 \\
2.2390 & 1.23e-12 & 4.54e-12 & 6.16e-13 & 4.39e-12 & 8.12e-15 & 1.58e-13 \\
2.3200 & 1.22e-12 & 4.48e-12 & 6.06e-13 & 4.24e-12 & 7.29e-15 & 1.43e-13 \\
2.3991 & 1.21e-12 & 4.46e-12 & 5.91e-13 & 4.12e-12 & 6.24e-15 & 1.26e-13 \\
2.4800 & 1.19e-12 & 4.44e-12 & 5.77e-13 & 3.99e-12 & 5.30e-15 & 1.11e-13 \\
2.5640 & 1.17e-12 & 4.18e-12 & 5.58e-13 & 3.64e-12 & 4.07e-15 & 8.62e-14 \\
2.6500 & 1.15e-12 & 4.25e-12 & 5.41e-13 & 3.57e-12 & 3.03e-15 & 7.22e-14 \\
2.7389 & 1.12e-12 & 4.12e-12 & 5.22e-13 & 3.34e-12 & 1.77e-15 & 5.07e-14 \\
2.8300 & 1.10e-12 & 3.98e-12 & 5.04e-13 & 3.10e-12 & 9.18e-16 & 3.53e-14 \\
2.9239 & 1.07e-12 & 3.84e-12 & 4.88e-13 & 2.93e-12 & 3.98e-16 & 2.37e-14 \\
3.0200 & 1.05e-12 & 3.69e-12 & 4.73e-13 & 2.75e-12 & 1.96e-16 & 1.59e-14 \\
3.1139 & 1.03e-12 & 3.57e-12 & 4.59e-13 & 2.62e-12 & 1.21e-16 & 1.18e-14 \\
3.2100 & 1.00e-12 & 3.44e-12 & 4.46e-13 & 2.49e-12 & 8.56e-17 & 8.86e-15 \\
3.3137 & 9.77e-13 & 3.32e-12 & 4.34e-13 & 2.39e-12 & 6.51e-17 & 7.26e-15 \\
\end{tabular}
\end{table*}
\begin{table*}
\begin{tabular}{lccccccc}
3.4200 & 9.52e-13 & 3.20e-12 & 4.23e-13 & 2.29e-12 & 5.19e-17 & 5.95e-15 \\
3.5287 & 9.29e-13 & 3.10e-12 & 4.13e-13 & 2.21e-12 & 4.29e-17 & 5.04e-15 \\
3.6400 & 9.06e-13 & 3.01e-12 & 4.03e-13 & 2.13e-12 & 3.63e-17 & 4.28e-15 \\
3.7536 & 8.85e-13 & 2.92e-12 & 3.95e-13 & 2.07e-12 & 3.12e-17 & 3.68e-15 \\
3.8700 & 8.64e-13 & 2.85e-12 & 3.87e-13 & 2.01e-12 & 2.70e-17 & 3.16e-15 \\
3.9886 & 8.46e-13 & 2.78e-12 & 3.81e-13 & 1.97e-12 & 2.36e-17 & 2.74e-15 \\
4.1100 & 8.27e-13 & 2.73e-12 & 3.75e-13 & 1.94e-12 & 2.07e-17 & 2.38e-15 \\
4.2335 & 8.12e-13 & 2.69e-12 & 3.72e-13 & 1.92e-12 & 1.82e-17 & 2.07e-15 \\
4.3600 & 7.97e-13 & 2.66e-12 & 3.68e-13 & 1.90e-12 & 1.60e-17 & 1.80e-15 \\
4.4885 & 7.85e-13 & 2.64e-12 & 3.67e-13 & 1.90e-12 & 1.40e-17 & 1.56e-15 \\
4.6200 & 7.73e-13 & 2.63e-12 & 3.65e-13 & 1.91e-12 & 1.22e-17 & 1.35e-15 \\
4.7534 & 7.65e-13 & 2.63e-12 & 3.66e-13 & 1.92e-12 & 1.07e-17 & 1.16e-15 \\
4.8900 & 7.57e-13 & 2.64e-12 & 3.66e-13 & 1.93e-12 & 9.42e-18 & 1.00e-15 \\
5.0333 & 7.49e-13 & 2.66e-12 & 3.67e-13 & 1.96e-12 & 8.23e-18 & 8.66e-16 \\
5.1800 & 7.41e-13 & 2.67e-12 & 3.67e-13 & 1.98e-12 & 7.16e-18 & 7.46e-16 \\
5.3331 & 7.29e-13 & 2.65e-12 & 3.63e-13 & 1.97e-12 & 6.16e-18 & 6.40e-16 \\
5.4900 & 7.17e-13 & 2.63e-12 & 3.59e-13 & 1.96e-12 & 5.24e-18 & 5.49e-16 \\
5.6481 & 6.65e-13 & 2.45e-12 & 3.28e-13 & 1.80e-12 & 4.38e-18 & 4.68e-16 \\
5.8100 & 6.09e-13 & 2.14e-12 & 2.97e-13 & 1.57e-12 & 3.54e-18 & 3.98e-16 \\
5.9730 & 4.14e-14 & 2.06e-13 & 3.16e-14 & 2.31e-13 & 2.73e-18 & 3.37e-16 \\
6.1400 & 1.76e-15 & 2.05e-14 & 2.65e-15 & 3.51e-14 & 1.96e-18 & 2.98e-16 \\
6.3129 & 4.70e-16 & 9.83e-15 & 7.95e-16 & 1.70e-14 & 1.33e-18 & 2.19e-16 \\
6.4900 & 2.12e-16 & 4.58e-15 & 3.53e-16 & 7.94e-15 & 9.50e-19 & 1.49e-16 \\
6.6728 & 1.33e-16 & 2.84e-15 & 2.17e-16 & 4.93e-15 & 6.94e-19 & 1.02e-16 \\
6.8600 & 9.61e-17 & 1.82e-15 & 1.54e-16 & 3.15e-15 & 5.17e-19 & 7.11e-17 \\
7.0526 & 7.54e-17 & 1.37e-15 & 1.20e-16 & 2.36e-15 & 3.88e-19 & 5.02e-17 \\
7.2500 & 6.17e-17 & 1.05e-15 & 9.68e-17 & 1.79e-15 & 2.93e-19 & 3.60e-17 \\
7.4476 & 5.22e-17 & 8.54e-16 & 8.11e-17 & 1.46e-15 & 2.23e-19 & 2.59e-17 \\
7.6500 & 4.50e-17 & 6.96e-16 & 6.94e-17 & 1.18e-15 & 1.70e-19 & 1.86e-17 \\
7.8575 & 3.93e-17 & 5.86e-16 & 6.02e-17 & 9.95e-16 & 1.30e-19 & 1.33e-17 \\
8.0700 & 3.48e-17 & 4.95e-16 & 5.29e-17 & 8.38e-16 & 1.01e-19 & 9.51e-18 \\
8.2923 & 3.09e-17 & 4.27e-16 & 4.66e-17 & 7.23e-16 & 7.79e-20 & 6.80e-18 \\
8.5200 & 2.75e-17 & 3.66e-16 & 4.13e-17 & 6.20e-16 & 6.10e-20 & 4.83e-18 \\
8.7522 & 2.47e-17 & 3.18e-16 & 3.67e-17 & 5.40e-16 & 4.83e-20 & 3.45e-18 \\
8.9900 & 2.21e-17 & 2.77e-16 & 3.28e-17 & 4.71e-16 & 3.89e-20 & 2.48e-18 \\
9.2321 & 1.99e-17 & 2.43e-16 & 2.93e-17 & 4.15e-16 & 3.19e-20 & 1.80e-18 \\
9.4800 & 1.79e-17 & 2.14e-16 & 2.62e-17 & 3.66e-16 & 2.65e-20 & 1.31e-18 \\
9.7320 & 1.61e-17 & 1.89e-16 & 2.34e-17 & 3.24e-16 & 2.25e-20 & 9.77e-19 \\
9.9900 & 1.45e-17 & 1.67e-16 & 2.09e-17 & 2.87e-16 & 1.94e-20 & 7.28e-19 \\
10.2421 & 1.30e-17 & 1.49e-16 & 1.87e-17 & 2.56e-16 & 1.70e-20 & 5.62e-19 \\
10.5000 & 1.17e-17 & 1.33e-16 & 1.67e-17 & 2.29e-16 & 1.50e-20 & 4.36e-19 \\
10.7962 & 1.03e-17 & 1.18e-16 & 1.46e-17 & 2.04e-16 & 1.31e-20 & 3.46e-19 \\
11.1000 & 9.02e-18 & 1.05e-16 & 1.27e-17 & 1.81e-16 & 1.14e-20 & 2.75e-19 \\
11.3964 & 7.87e-18 & 9.31e-17 & 1.10e-17 & 1.61e-16 & 9.88e-21 & 2.24e-19 \\
11.7000 & 6.79e-18 & 8.28e-17 & 9.42e-18 & 1.44e-16 & 8.54e-21 & 1.82e-19 \\
11.9965 & 5.82e-18 & 7.41e-17 & 8.04e-18 & 1.28e-16 & 7.37e-21 & 1.51e-19 \\
12.3000 & 4.92e-18 & 6.64e-17 & 6.75e-18 & 1.15e-16 & 6.28e-21 & 1.26e-19 \\
12.6455 & 3.99e-18 & 5.92e-17 & 5.44e-18 & 1.03e-16 & 5.15e-21 & 1.05e-19 \\
13.0000 & 3.12e-18 & 5.28e-17 & 4.22e-18 & 9.16e-17 & 4.06e-21 & 8.81e-20 \\
13.3457 & 2.35e-18 & 4.76e-17 & 3.16e-18 & 8.26e-17 & 3.09e-21 & 7.41e-20 \\
13.7000 & 1.66e-18 & 4.33e-17 & 2.21e-18 & 7.51e-17 & 2.20e-21 & 6.27e-20 \\
14.0459 & 1.07e-18 & 4.01e-17 & 1.42e-18 & 6.94e-17 & 1.43e-21 & 5.36e-20 \\
14.4000 & 5.75e-19 & 3.82e-17 & 7.56e-19 & 6.60e-17 & 7.71e-22 & 4.72e-20 \\
14.7461 & 2.08e-19 & 3.78e-17 & 2.71e-19 & 6.53e-17 & 2.79e-22 & 4.18e-20 \\
15.1000 & 1.80e-23 & 1.20e-17 & 2.36e-23 & 2.07e-17 & 4.86e-26 & 1.19e-20 \\
\hline
\end{tabular}
\caption{Equivalent-equilibrium photoionization and photoheating rates for \hbox{H~$\scriptstyle\rm I$}, \HeI and \HeII as a function of redshift. Note that these rates should only be used for following photoheating and photoionization in cosmological simulations with codes that assume ionization equilibrium. Electronic versions of this table are available both on the journal and arXiv article web pages. For other purposes or for codes that use a full non-equilibrium ionization and heating solver, use the rates provided in Table~\ref{tab:rates}.}
\label{tab:eq_rates}
\end{table*}

\end{document}